\documentclass[twocolumn]{revtex4}

\usepackage[usenames,dvipsnames]{color}

\usepackage[latin1]{inputenc}
\usepackage{graphicx}
\usepackage{amssymb}
\usepackage{subfigure}
\usepackage{stackrel}

\usepackage{amsmath}
\usepackage{amsfonts}
\usepackage{bm}
\usepackage{color}
\usepackage{verbatim} 

\newcommand{\ket}[1]{\ensuremath{\left|{#1}\right\rangle}}
\newcommand{\bra}[1]{\ensuremath{\left\langle{#1}\right |}}

\newcommand{\beq}{\begin{equation}}
\newcommand{\eeq}{\end{equation}}
\newcommand{\bse}{\begin{subequations}}
\newcommand{\ese}{\end{subequations}}
\newcommand{\bea}{\begin{eqnarray}}
\newcommand{\eea}{\end{eqnarray}}
\newcommand{\bit}{\begin{itemize}}
\newcommand{\eit}{\end{itemize}}
\newcommand{\bpmatrix}{\begin{pmatrix}}
\newcommand{\epmatrix}{\end{pmatrix}}

\newcommand{\be}{\begin{equation}}
\newcommand{\ee}{\end{equation}}
\newcommand{\ben}{\begin{eqnarray}}
\newcommand{\een}{\end{eqnarray}}

\newcommand{á}{\'a}
\newcommand{é}{\'e}
\newcommand{í}{\'i}
\newcommand{ó}{\'o}
\newcommand{ú}{\'u}

\begin{document}

\title{Guaranteed emergence of genuine entanglement in 3-qubit evolving systems}

\author{A. Valdés-Hernández, V. T. Brauer, F. S. Zamora}
\affiliation{Instituto de F\'{\i}sica, Universidad Nacional Aut\'{o}noma de M\'{e}xico, \\
Apartado Postal 20-364, Ciudad de México, Mexico}

\email[]{andreavh@fisica.unam.mx}

\begin{abstract}
Multipartite entanglement has been shown to be of particular relevance for a better understanding and exploitation of the dynamics and flow of entanglement in multiparty systems. This calls for analysis aimed at identifying the appropriate processes that guarantee the emergence of multipartite entanglement in a wide range of scenarios. Here we carry on such analysis considering a system of two initially entangled qubits, one of which is let to interact with a third qubit according to an arbitrary unitary evolution. We establish necessary and sufficient conditions on the corresponding Kraus operators, to discern whether the evolved state pertains to either one of the classes of 3-qubit pure states that exhibit some kind of entanglement, namely biseparable, W-, and GHZ- genuine entangled classes. Our results provide a classification of the Kraus operators according to their capacity of producing 3-qubit entanglement, and pave the way for extending the analysis to larger systems and determining the particular interactions that must be implemented in order to create, enhance and distribute entanglement in a specific manner.
\end{abstract}

\maketitle

\section{Introduction}

Quantum entanglement constitutes an extremely rich phenomenon that has occupied a 
prominent role in contemporary physics. Ranging from inquiries regarding foundational
aspects of quantum mechanics, to the search of methods aimed at efficiently exploiting 
this resource in quantum information tasks, the lines of research involving entanglement 
have been rapidly multiplied in the last decades. One of them, which is of particular 
interest for the present analysis, focuses on the dynamics and flow of entanglement in multiparty systems. 

The study of the evolution of initially entangled systems has made clear that entanglement is a dynamical resource that is considerable enriched in 
multipartite systems \cite{Horo,multi,AolDyn}. In particular, multipartite entanglement in qubit systems has been shown to play a relevant role in the construction of dynamical invariants of entanglement that allow to introduce the notion of a flow of correlations among the constituents of the complete system \cite{14,16,17}. The analysis of the distribution of entanglement acquires special relevance in the context of open quantum systems, where decoherence processes are involved \cite{AolDyn,25}. Whereas the interaction of an entangled bipartite system with an environment tends to degrade their entanglement, the inclusion of the environment transforms the bipartite system into a multiparty one, thus allowing for the 
entanglement to distribute in the form of bipartite and multipartite entanglement between the different subsystems. Therefore, in appropriate circumstances the disentanglement of some subsystems is accompanied by the entanglement of others, giving rise to phenomena such as the transference and transformation of entanglement assisted by the environment. This observations have allowed, for example, to understand phenomena such as the Sudden Death of Entanglement \cite{12,13,12a}, characterized by the fact that the extinction of the entanglement precedes the extinction of the coherences.  

Further, the analysis of the dynamics of entangled systems has posed questions on the fragility of this resource in different kind of entangled states, on the development of strategies to increase some type of entanglement in decoherence processes, and more ambitiously on the establishment of the global laws that govern the evolution of entanglement in multiparty systems. To take a step further in these directions it becomes necessary to identify the appropriate processes that guarantee the emergence of specific classes of entanglement, particularly of multipartite entanglement. Here we tackle this problem for a 3-qubit pure state. In spite of being the simplest multiparty system, it is well-known that it allows for the existence of 3-partite entangled states that have been successfully realized in different physical systems \cite{re-1,re-2,re-3,re-4}, and exhibit different robustness and features \cite{23} that are required in various quantum information tasks \cite{qi1,qi2,qi3,qi4}. 

We contribute to the study of the emergence of genuine tripartite entanglement by considering a system of two initially entangled qubits, one of which is let to interact with a third qubit under an \emph{arbitrary} unitary transformation. Resorting to the formalism of Kraus operators \cite{AolDyn,Kraus} to represent the evolution, we classify them according to their capacity of producing genuine tripartite entanglement, either of the W- or GHZ-type, which are the two inequivalent classes of multipartite entanglement in 3-qubit systems \cite{22,23,EntDet2009}. We establish necessary and sufficient conditions ---based on basic and similarity-invariant properties of the Kraus operators--- for the evolved state to pertain to each of the three families of 3-qubit pure states that exhibit some kind of entanglement, namely biseparable, W-, and GHZ- entangled classes. Our results enrich and add to other classification schemes aimed at distinguishing the particular nature of the entanglement of a 3-qubit state \cite{d3-2,d3-1,Datta18,DasBell,Zhao13,Brunner12}, by offering a classification of the \emph{evolution operators} that drive the initial state into one with a specific type of entanglement. Moreover, by relating the emergence of 3-partite correlations with the structure of the Kraus operators, our classification settles the basis for establishing the type of operations (or Hamiltonians) that must be implemented in order to create and distribute entanglement in a convenient way.
 
The paper is structured as follows. Sections \ref{Kraus} and \ref{3p} contain preliminary material related, respectively, to the Kraus formalism applied to the system of interest, and to the emergence of genuine tripartite entanglement. In Section \ref{GHZ} we find an expression for the so-called 3-tangle \cite{19} in terms of the initial entanglement and an invariant function of the Kraus operators, and in Section \ref{W} we derive the expressions for \emph{all} the possible bipartite entanglements also in terms of the Kraus operators. We then present our main results in Section \ref{main}, where necessary and sufficient conditions to guarantee the different kinds of entanglement in the evolved state are established, in terms of the Kraus operators. We discuss the dependence of the generation of tripartite entanglement on the amount of initial bipartite entanglement, and look briefly into the extension of our results to 4-qubit systems evolving under local channels. In Section \ref{ex} the emergence and distribution of entanglement is analyzed for some examples involving paradigmatic quantum decoherence maps, and some final comments are presented in Section \ref{conclu}.

\section{Kraus operators and quantum maps}\label{Kraus}
Consider a 3-qubit system composed of subsystems $S$, $S'$, and $E$. The tripartite system is assumed to be initially in a pure state of the form
\begin{equation}\label{OriginalState}
\ket{\phi(0)}_{S'SE} = \ket{\psi_0}_{S'S}\ket{0}_{E},
\end{equation}
with $\ket{\psi_0}_{S'S}$ an arbitrary state of the $S+S'$ subsystem.
$S$ then starts to interact with $E$, so the evolution operator decomposes as $U(p)=\mathbb I_{S'}\otimes U_{SE}(p)$, with $U_{SE}$ an arbitrary unitary transformation, and $p$ a real parameter (e.g., an appropriate time parametrization) such that at $p=0$, $U_{SE}(0)=\mathbb{I}_{SE}$. The evolved state has the general form (with $n,l,m = 0,1$)
\begin{equation}\label{Evolvedstate}
\ket{\phi(p)}_{S'SE} = \sum_{nlm}c_{nlm}(p)\ket{nlm}_{S'SE}, 
\end{equation}
with specific coefficients $c_{nlm}$ that will be determined below. The corresponding tripartite density matrix reads 
\begin{eqnarray}
\rho(p)&=&\ket{\phi(p)}\bra{\phi(p)}=U(p)\rho(0) U^{\dagger}(p)\\ \label{rho}
&=&[\mathbb I_{S'}\otimes U_{SE}(p)](\ket{\psi_0}\bra{\psi_0}\otimes\ket{0}\bra{0}) [\mathbb I_{S'}\otimes U^{\dagger}_{SE}(p)], \nonumber
\end{eqnarray}
and the (reduced) state of subsystem $S$ follows a (in general non-unitary) evolution given by \cite{AolDyn,Kraus}
\begin{equation}\label{rhoS}
\rho_{S}(p) = \textrm{Tr}_{(S'E)}\rho(p)=\sum_{\mu=0,1}K_{\mu}(p)\rho_0K_{\mu}^{\dagger}(p),
\end{equation}
where $\rho_0\equiv\rho_S(0)$, and $\{K_\mu(p) \equiv \bra{\mu}_EU_{SE}(p)\ket{0}_E\}$ stand for the Kraus operators associated to the transformation $U_{SE}$, which satisfy the restriction
\beq \label{sumK}
\sum_{\mu} K_{\mu}^{\dagger}K_{\mu} = \mathbb{I}_S
\eeq 
that guarantees the normalization condition $\textrm{Tr}\rho_S(p) = 1$. 

In terms of the Kraus operators, the unitary evolution of the $S+E$ system can be alternatively represented by the following quantum map:
\begin{eqnarray} \label{map}
\ket{00}_{SE} &\xrightarrow[U_{SE}]{}& (K_0(p)\ket{0}_S)\ket{0}_E+(K_1(p)\ket{0}_S)\ket{1}_E, \\
\ket{10}_{SE} &\xrightarrow[U_{SE}]{}& (K_0(p)\ket{1}_S)\ket{0}_E+(K_1(p)\ket{1}_S)\ket{1}_E. \nonumber
\end{eqnarray}
By writing the matrix representation of $K_0$ and $K_1$ in the basis $\{\ket{0}=(1,0)^\top, \ket{1}=(0,1)^\top\}$ as
 \begin{equation}
  K_0 = \left( \begin{array}{cc}
m_{00} & m_{01}  \\
m_{10} & m_{11}  \\
 \end{array} \right), K_1 = \left( \begin{array}{cc}
n_{00} & n_{01}  \\
n_{10} & n_{11}  \\
 \end{array} \right),
 \end{equation}
with $m_{ij}=m_{ij}(p)$, and $n_{ij}=n_{ij}(p)$, the transformation (\ref{map}) can be recasted in the form 
\begin{eqnarray}
\label{mapa}
\!\!\!\!\!\ket{00}_{SE} \!\!&\xrightarrow[U_{SE}]{}&\!\!m_{00}\ket{00} + m_{10}\ket{10}+ n_{00}\ket{01} + n_{10}\ket{11}\!,  \\
\!\!\!\!\!\ket{10}_{SE} \!\!&\xrightarrow[U_{SE}]{}&\!\!m_{01}\ket{00}+ m_{11}\ket{10}
  +n_{01}\ket{01}+ n_{11}\ket{11}\!.\nonumber 
\end{eqnarray}
The matrix elements of the Kraus operators represent thus the probability amplitudes of the transformed states, and therefore can be used to describe the effective evolution associated to the unitary transformation. In what follows we will resort to this (Kraus) formalism to extract conclusions regarding the dynamics of entanglement, via an analysis of the properties of the matrices $K_{\mu}$ (in what follows we omit the explicit dependence of these matrices on the parameter $p$).
 
\section{Emergence of genuine entanglement}\label{3p}
Since the tripartite state $\rho(p)$ is pure, a suitable measure of the (bipartite) entanglement between subsystems $i$ and $j+k$ (with $i,j,k\in\{S,S',E\}$) is given by the tangle, or squared concurrence $C^2$, given by \cite{18},
\begin{equation}\label{Concurrencia}
C^2_{i|jk} = 2(1-\textrm{Tr}\rho_i^2)=2(1-\textrm{Tr}\rho_{jk}^2), 
\end{equation}
where $\rho_i=\textrm{Tr}_{(jk)}\rho$ stands for the reduced density matrix of subsystem $i$. The information regarding the distribution of entanglement among the different constituents of the system is contained in the following (CKW) decomposition \cite{19} 
\begin{equation} \label{Concurrencia biparticiones}
C^2_{i|jk}(p) = C^2_{ij}(p) + C^2_{ik}(p)+\tau(p),
\end{equation}
which involves the concurrence  $C_{ij}$ ---that measures the entanglement between subsystems $i$ and $j$ \cite{concu}--- and the so-called 3-tangle, $\tau$. For a general tripartite state of the form (\ref{Evolvedstate}), and in terms of $a_{ij}\equiv c_{0ij}$ and $b_{ij}\equiv c_{1ij}$, it is given by \cite{19}
\begin{equation}
\tau =4\vert d_{1}-2d_{2}+4d_{3}\vert, \label{tau}
\end{equation}
where
\begin{subequations}
\begin{eqnarray}
d_{1}&=& a^{2}_{00}b^{2}_{11}+a^{2}_{01}b^{2}_{10}+a^{2}_{10}b^{2}_{01}+a^{2}_{11}b^{2}_{00},\\
d_{2}&=&a_{00}a_{11}b_{00}b_{11}+a_{01}a_{10}b_{10}b_{01}+\\ &&(a_{10}b_{01}+a_{01}b_{10})(a_{00}b_{11}+a_{11}b_{00}),\nonumber\\
d_{3}&=& a_{00}a_{11}b_{10}b_{01}+a_{01}a_{10}b_{00}b_{11}\label{d3}.
\end{eqnarray}
\end{subequations}

The quantity $\tau$ divides the set of 3-qubit pure states into two inequivalent (under Stochastic Local 
Operations and Classical Communication) families \cite{22}: those for which $\tau\neq0$ (GHZ-family), and 
those for which $\tau=0$ (W-family). In the former case, the nonzero 3-tangle guarantees that the states in the GHZ-family are 
entangled in \emph{all} bipartitions, meaning that they are genuine tripartite entangled \cite{EntDet2009}. In its 
turn, the W-family contains three (also inequivalent) subfamilies, comprising: i) fully separable states, for 
which $C^2_{i|jk}=0$ for \emph{all} $i$, hence no entanglement is present; ii) biseparable states, for which 
$C^2_{i|jk}=0$ for a \emph{single} subsystem $i$, so that (only) one subsystem is disentangled from the rest; 
and iii) 3-partite entangled states, satisfying $C^2_{i|jk}>0$ for \emph{all} $i$, hence the state is genuine 
tripartite entangled. The members of this last (sub)family constitute 3-partite entangled states whose 
multipartite entanglement is not detected by $\tau$. Therefore, $\tau$ is considered as a quantitative 
measure of (only) GHZ-type genuine entanglement.

The lack of a suitable measure of W-type genuine entanglement difficults to determine (without 
explicitly calculating all $C^2_{i|jk}$) whether a given state is genuine tripartite entangled, 
biseparable, or fully separable. However, some classification criteria have been established, based on: entangled hypergraphs \cite{d3-2}; hyper- and sub-determinants, providing 
necessary and sufficient conditions for separability \cite{d3-1}; construction of appropriate observables 
\cite{Datta18} and Bell inequalities \cite{DasBell} that determine the family to which a given state pertains; 
expectation values of Pauli operators that distinguishes between fully separable, biseparable and 3-partite entangled states \cite{Zhao13}, and device-independent witnesses that discriminate between W- and GHZ-type entanglement \cite{Brunner12}. It is the aim of the following sections to enrich the existent criteria from a dynamical perspective, establishing the conditions on the Kraus operators, according to their capacity of producing each type of entanglement in the evolved state $\ket{\phi(p)}$.  
\section{3-tangle in terms of the Kraus operators}\label{GHZ}
Let us introduce the matrices 
\begin{equation}
C_0=[a_{ij}]=[c_{0ij}],\quad C_1=[b_{ij}]=[c_{1ij}],
\end{equation}
and rewrite Eq. $(\ref{d3})$ as
\begin{eqnarray}
d_{3}&=&a_{00}a_{11}(b_{10}b_{01}-b_{00}b_{11})+a_{01}a_{10}(b_{00}b_{11}-b_{10}b_{01})+\nonumber\\
&&+a_{11}a_{00}b_{00}b_{11}+a_{01}a_{10}b_{01}b_{10}\nonumber\\
&=&\!\!-\det (C_0C_1)+a_{11}a_{00}b_{00}b_{11}+a_{01}a_{10}b_{01}b_{10}.
\end{eqnarray}
Direct calculation thus gives
\begin{eqnarray}
d_{1}-2d_{2}+4d_{3}&=&-4\det (C_{0}C_{1})+\\
&&+(a_{00}b_{11}+a_{11}b_{00}-a_{01}b_{10}-a_{10}b_{01})^{2}.\nonumber
\end{eqnarray}
Recognizing in the second line of this expression the quantity
\begin{gather}
\textrm{Tr}\,C_{0}\,\textrm{Tr}\,C_{1}-\textrm{Tr}\,(C_{0}C_{1})=\\
=a_{00}b_{11}+a_{11}b_{00}-a_{01}b_{10}-a_{10}b_{01},\nonumber
\end{gather}
we obtain the following simplified expression of Eq. (\ref{tau}):
\begin{equation}\label{tauB}
\tau=4\,|4\det (C_{0}C_{1})-g^2(C_{0},C_{1})|,
\end{equation}
where
\begin{equation}\label{g}
g(A,B)\equiv\textrm{Tr}A\,\textrm{Tr}B-\textrm{Tr}\,(AB),
\end{equation}
so that
\begin{subequations}\label{propsg}
\begin{gather}
g(A,B)=g(B,A)=g^{*}(A^{\dagger},B^{\dagger}),\\
g(\lambda_1A,\lambda_2\,[B+C])=\lambda_1\lambda_2\,[\,g(A,B)+g(A,C)],
\end{gather}
\end{subequations}
with $\lambda_{1,2}$ arbitrary complex numbers.

Notice that $g(A,B)$ is a similarity-invariant function, and that whenever $A$ and $B$ are $2\times2$ matrices, Eq. (\ref{g}) is equivalent to
\begin{equation}
g(A,B)=\det(A+B)-\det A-\det B,\label{gdet}
\end{equation}
and thus the following property holds:
\begin{eqnarray}
g(A,B)g(C,D)\!\!&=&\!\!g(AC,BD)+g(AD,BC).\label{gprod}
\end{eqnarray}
In particular, for $A=B$ this gives $g(A,A)g(C,D)=2\,g(AC,AD)$, and with the aid of Eq. (\ref{gdet}) we get
\begin{eqnarray}
g(AC,AD)=\det A \, g(C,D).\label{gprod2}
\end{eqnarray}
Also, from Eqs. (\ref{g}) and (\ref{gdet}) it follows that for $A$ with unit trace, $1-\textrm{Tr}A^2=2\det A$, whence
\begin{equation}\label{deti}
C^2_{i|jk}= 2(1-\textrm{Tr}\rho_i^2)=4\det \rho_i.
\end{equation}

Let us now come back to the initial state (\ref{OriginalState}), and write $\ket{\psi_0}_{S'S}$ in its general form:
\beq \label{psi0}
\ket{\psi_0}_{S'S}=\alpha\ket{11} + \beta\ket{10} +\gamma\ket{01}+\delta\ket{00},
\eeq
with $|\alpha|^2+|\beta|^2 + |\gamma|^2 + |\delta|^2 = 1$. Under the map (\ref{mapa}), the coefficients of the evolved state (\ref{Evolvedstate}) are such that 
\begin{subequations}\label{Cs}
\begin{eqnarray}
C_0&=&K_0M_{0}(\delta,\gamma)+K_1M_{1}(\delta,\gamma),\\
 C_1&=&K_0M_{0}(\beta,\alpha)+K_1M_{1}(\beta,\alpha),
\end{eqnarray}
\end{subequations}
where $M_{0,1}(x,y)$ are the matrices
\begin{equation}
M_{0}(x,y)=\left( \begin{array}{cc}
x & 0  \\
y & 0  \\
 \end{array} \right),\quad M_{1}(x,y)=\left( \begin{array}{cc}
0 & x  \\
0 & y  \\
 \end{array} \right).
\end{equation}
Substitution of Eqs. (\ref{Cs}) into (\ref{tauB}) leads, after some algebraic manipulation, to
\begin{equation}\label{FenDet}
\tau=\mathcal{E}^2_0\big|4\det (K_0K_1) -g^2(K_0,K_1)\big|,
\end{equation}
where $\mathcal{E}_0$ stands for the initial entanglement between $S$ and $S'$ (hereafter assumed to be nonzero):
\begin{equation}\label{E0detrhoB}
\mathcal{E}^2_0\equiv C^2_{S'S}(0)=C^2_{S'|SE}(0)=C^2_{S|S'E}(0)=4\det \rho_0.
\end{equation}

Equation (\ref{FenDet}) shows that the amount of 3-tangle depends on the initial state only through the initial (bipartite) entanglement, and that its dynamics is completely determined by the Kraus operators that characterize the quantum map (\ref{map}). As discussed in \cite{16} (where a similar, though less simplified, version of Eq. (\ref{FenDet}) was derived), the 3-tangle thus emerges as a mere redistribution of the initial entanglement $C^2_{S'S}(0)$, induced by the interaction between $S$ and $E$. 

\section{Bipartite entanglement in terms of the Kraus operators}\label{W}
Now that we have the expression (\ref{FenDet}) for the 3-tangle as a function of the matrices $\{K_{\mu}\}$ that determine the dynamics of the system, our next aim is to find the expressions of the bipartite entanglements $C^2_{i|jk}$ and $C^2_{ij}$, also in terms of the Kraus operators. 

Since the evolved state $\rho(p)$ results from applying a unitary transformation in the $S+E$ subsystem, the entanglement in the partition $S'|SE$ is not affected at all during the evolution. Consequently 
\beq \label{S'}
C^2_{S'|SE}(p)=C^2_{S'|SE}(0)=\mathcal{E}^2_0,
\eeq
and $C^2_{S'|SE}$ is independent of the specific Kraus operators. In what follows we thus focus on the entanglement $C^2_{i|jk}$ for $i=S,E$. 

Resorting to Eqs. (\ref{rhoS}) and (\ref{deti}) we get, with the aid of (\ref{gdet}),
\begin{eqnarray}\label{S}
C^2_{S|S'E}&=&4\det \big(K_{0}\rho_0K_{0}^{\dagger}+K_{1}\rho_0K_{1}^{\dagger}\big)\nonumber\\
&=&4\det \rho_0\big(|\det K_0|^2+|\det K_1|^2\big)+\nonumber\\
 &&+\,4\,g\big(K_0\rho_0K^{\dagger}_0,K_1\rho_0K^{\dagger}_1\big).
\end{eqnarray}
Now, from Eqs. (\ref{gprod}) and (\ref{gprod2}) we obtain
\begin{gather}
g\big(K_0\rho_0K^{\dagger}_0,K_1\rho_0K^{\dagger}_1\big)=\nonumber \\
=\det \rho_0\big|g\big(K_0,K_1\big)\big|^2-
g\big(K_0\rho_0K^{\dagger}_1,K_1\rho_0K^{\dagger}_0\big),
\end{gather}
and therefore
\begin{eqnarray}\label{S2a}
C^2_{S|S'E}&=&\mathcal{E}^2_0\big(|\det K_0|^2+|\det K_1|^2+|g^{2}(K_0,K_1)|\big)-\nonumber\\
 &&-\,4\,g(K_0\rho_0K^{\dagger}_1,K_1\rho_0K^{\dagger}_0).
\end{eqnarray}

As for $C^2_{E|SS'}$, we start from Eq. (\ref{rho}) and write (with $\{\ket{n}\}$ and $\{\ket{n'}\}$ orthonormal basis of the Hilbert space of $S$ and $S'$, respectively, and $\{\ket{\mu}\}$, $\{\ket{\nu}\}$ orthonormal basis of the Hilbert space of $E$)
\begin{eqnarray}
\rho_E&=&\textrm{Tr}_{(SS')}\rho=\sum_{nn'}\bra{nn'}U(\ket{\psi_0}\bra{\psi_0}\otimes \ket{0}\bra{0})U^{\dagger}\ket{nn'}\nonumber\\
&=&\sum_{n}\bra{n}U_{SE}\ket{0}\Big(\sum_{n'}\langle{n'}|\psi_0\rangle\langle \psi_0|n'\rangle\Big)\bra{0}U^{\dagger}_{SE}\ket{n}\nonumber\\
&=&\sum_{n\mu\nu}\bra{n}\ket{\mu}\bra{\mu}U_{SE}\ket{0}\rho_0\bra{0}U^{\dagger}_{SE}\ket{\nu}\bra{\nu}\ket{n}\nonumber\\
&=&\sum_{\mu\nu}\Big(\sum_{n}\bra{n}K_{\mu}\rho_0K^{\dagger}_{\nu}\ket{n}\Big)\ket{\mu}\bra{\nu}\nonumber\\
&=&\sum_{\mu\nu}\textrm{Tr}\,\big(K_{\mu}\rho_0K^{\dagger}_{\nu}\big)\ket{\mu}\bra{\nu}.\label{rhoE}
\end{eqnarray}
Equation (\ref{deti}) thus leads to (using the cyclic property of the trace)
\begin{eqnarray}\label{E}
C^2_{E|SS'}&=&4\big[\textrm{Tr}\,\big(\rho_0K^{\dagger}_{0}K_{0}\big)
\textrm{Tr}\,\big(\rho_0K^{\dagger}_{1}K_{1}\big)-\nonumber\\
&&-\textrm{Tr}\,\big(K_{0}\rho_0K^{\dagger}_{1}\big)
\textrm{Tr}\,\big(K_{1}\rho_0K^{\dagger}_{0}\big)
\big].
\end{eqnarray}
We now resort to Eq. (\ref{g}) to get
\begin{gather}\label{E2}
C^2_{E|SS'}=\nonumber\\=4\big[g(\rho_0K^{\dagger}_{0}K_{0},\rho_0K^{\dagger}_{1}K_{1}\big)+\textrm{Tr}\,\big(\rho_0K^{\dagger}_{0}K_{0}\,\rho_0K^{\dagger}_{1}K_{1}\big)-\nonumber\\
-g(K_{0}\rho_0K^{\dagger}_{1},K_{1}\rho_0K^{\dagger}_{0}\big)-\textrm{Tr}\,\big(K_{0}\rho_0K^{\dagger}_{1}K_{1}\rho_0K^{\dagger}_{0}\big)
\big].
\end{gather}
The trace terms in this expression cancel each other due to the cyclic property of the trace, and Eq. (\ref{gprod2}) leads finally to
\begin{eqnarray}
C^2_{E|SS'}&=&\mathcal{E}^2_0\, g\big(K^{\dagger}_{0}K_{0},K^{\dagger}_{1}K_{1}\big)-\nonumber\\
&&-4\,g(K_{0}\rho_0K^{\dagger}_{1},K_{1}\rho_0K^{\dagger}_{0}\big).\label{E3}
\end{eqnarray}

Gathering results we get:
\begin{subequations}\label{SS'E}
\begin{eqnarray}
C^2_{S'|SE}&=&\mathcal{E}^2_0,\\
C^2_{S|S'E}&=&\mathcal{E}^2_0D_S(K)+G(K,\rho_0),\label{Sbis}\\
C^2_{E|SS'}&=&\mathcal{E}^2_0D_E(K)+G(K,\rho_0),\label{Ebis}
\end{eqnarray}
\end{subequations}
where
\begin{subequations}\label{D}
\begin{eqnarray}
\!\!\!\!\!\!\!\!D_S(K)\!\!&=&|\det K_0|^2+|\det K_1|^2+|g^2(K_0,K_1)|,\\
\!\!\!\!\!\!\!\!D_E(K)\!\!&=&\!\!g\big(K^{\dagger}_{0}K_{0},K^{\dagger}_{1}K_{1}\big),
\end{eqnarray}
\end{subequations}
represent contributions independent of the initial conditions, and
\begin{eqnarray}\label{G}
G(K,\rho_0)=-4\,g(K_{0}\rho_0K^{\dagger}_{1},K_{1}\rho_0K^{\dagger}_{0}\big)
\end{eqnarray}
represents the contribution (common to $C^2_{S|S'E}$ and $C^2_{E|SS'}$) that bears the information regarding the initial state $\rho_0$. Now, resorting to the condition (\ref{sumK}) satisfied by the Kraus operators we obtain $g\big(K^{\dagger}_{0}K_{0},K^{\dagger}_{1}K_{1}\big)=1-|\det K_0|^2-|\det K_1|^2$, a result that relates $D_S$ and $D_E$ according to
\begin{equation}\label{DEDS}
D_S=1-D_{E}+|g^2(K_0,K_1)|=1+g\big(K^{\dagger}_{0}K_{1},K^{\dagger}_{1}K_{0}\big).
\end{equation}

From the CKW decomposition (\ref{Concurrencia biparticiones}) it follows that
\begin{equation}
 C^2_{ij}=\tfrac{1}{2}(C^2_{i|jk}+C^2_{j|ik}-C^2_{k|ij}-\tau),
\end{equation}
whence Eqs. (\ref{FenDet}), (\ref{SS'E}), and (\ref{DEDS}) give
\begin{subequations}\label{bip}
\begin{align}
C^{2}_{S'S}&=\mathcal{E}^2_0\big(\vert \det K_{0}\vert + \vert \det K_{1}\vert\big)^{2} -\nonumber\\
&- \tfrac{1}{2}\mathcal{E}^2_0\big(|u|-|v|+|u-v|\big), \label{SS'}
\\
C^{2}_{S'E}&=\mathcal{E}^2_0\big[1-\big(\vert \det K_{0}\vert + \vert \det K_{1}\vert\big)^{2}\big]-\nonumber\\
&-\tfrac{1}{2}\mathcal{E}^2_0\big(|v|-|u|+|u-v|\big),\label{S'E}\\
C^{2}_{SE}&=G+\tfrac{1}{2}\mathcal{E}^2_0\big(|v|-|u-v|\big),\label{SE}
\end{align}
\end{subequations}
where we wrote
\begin{equation}
u=4\det (K_{0}K_{1}),\quad v=g^2(K_{0},K_{1}),
\end{equation} 
so that $\tau=\mathcal{E}^2_0|u-v|$. Notice from Eq. (\ref{SE}) that whenever $u$ vanishes (at least one $K_{\mu}$ has vanishing determinant), $G$ coincides with $C^2_{SE}$. In such case the contribution $G(K,\rho_0)$ in Eqs. (\ref{Sbis}) and (\ref{Ebis}) represents the entanglement directly \emph{generated} as a result of the interaction between $S$ and $E$, and consequently $D_{S}$ and $D_E$ represent the amount of entanglement that is being \emph{distributed}, from the initial available entanglement $\mathcal{E}^2_0$.

Resorting to the inequality $|u-v|\leq |u|+|v|$ we find the following lower bounds of $C^2_{ij}$: 
\begin{subequations}\label{cotas}
\begin{eqnarray}
&\mathcal{E}^2_0\big(\vert \det K_{0}\vert - \vert \det K_{1}\vert\big)^{2}\leq C^{2}_{S'S}, \label{SS'c}
\\
&\!\!\!\!\!\!\!\!\!\mathcal{E}^2_0\!\big[1\!-\!\big(\vert \!\det K_{0}\vert \!+\! \vert \!\det K_{1}\vert\big)^{2}\!\!-\!|g^2(K_{0},K_{1})|\big]\!\!\leq \!C^{2}_{S'E},\label{S'Ec}\\
&G-2\mathcal{E}^2_0|\det (K_{0}K_{1})|\leq C^{2}_{SE}.\label{SEc}
\end{eqnarray}
\end{subequations}
The equality sign holds whenever $|u-v|=|u|+|v|$. This will occur, in particular, whenever $u=0$ and/or $v=0$, which is the case of all the examples below. 

\section{Necessary and sufficient conditions for the emergence of genuine entanglement}\label{main}
Our aim now is to establish the conditions under which the transition from bipartite to multipartite entanglement is guaranteed. Specifically, we focus on the general properties that the Kraus operators able to create tripartite entanglement must satisfy. 

We first resort to Eq. (\ref{FenDet}),  
which allows us to determine whether the evolved state $\ket{\phi(p)}$ exhibits GHZ-type genuine entanglement ($\tau\neq0$), or not ($\tau=0$), as follows:
\begin{equation} \label{iff}
4\det (K_0K_1)\!\neq \!g^2(K_0,K_1)\!\! \iff \!\! \ket{\phi(p)} \textrm{has 3-tangle.} 
\end{equation}
From here it follows, in particular, that if the quantities involved are real, then $\det (K_0K_1) < 0$ is a sufficient condition to produce 3-tangle.  

The result (\ref{iff}) establishes necessary and sufficient conditions (on the Kraus operators) for the state (\ref{Evolvedstate}) to pertain to the GHZ-family, yet it provides only necessary conditions for the state to belong to the W-(sub)family of 3-partite entangled states. In order to determine sufficient conditions for the emergence of W-type multipartite entanglement, we must assure that, while having $\tau=0$, the state is entangled in all bipartitions, i.e., $C^2_{i|jk}=C^2_{ij}+C^2_{ik}>0$ for all $i=S,S', E$. As seen from Eq. (\ref{S'}), this is automatically met for $i=S'$ (recall that we assume nonzero $\mathcal{E}_0$). 

Now, since $\tau=0$ we have $4\det (K_0K_1)=g^2(K_0,K_1)$, or $u=v$, whence Eqs. (\ref{bip}) reduce to 
\begin{subequations}\label{bip2}
\begin{align}
C^{2}_{S'S}&=\mathcal{E}^2_0\big(\vert \det K_{0}\vert + \vert \det K_{1}\vert\big)^{2} , \label{SS'tau0}
\\
C^{2}_{S'E}&=\mathcal{E}^2_0-C^{2}_{S'S}\label{S'Etau0},
\\
C^{2}_{SE}&=G+2\mathcal{E}^2_0|\!\det (K_{0}K_{1})|,\label{SEtau0}
\end{align}
\end{subequations}
and therefore
\begin{subequations}\label{SS'Ebis}
\begin{eqnarray}
C^2_{S|S'E}&=&\mathcal{E}^2_0d_S(K)+C^2_{ES},\label{Sbisb}\\
C^2_{E|SS'}&=&\mathcal{E}^2_0d_E(K)+C^2_{ES},\label{Ebisb}
\end{eqnarray}
\end{subequations}
where $\mathcal{E}^2_0d_S=C^2_{SS'}\geq0$, $\mathcal{E}^2_0d_E=C^2_{ES'}\geq0$, and 
\begin{equation}\label{Db}
d_S=1-d_E=\big(|\det K_0|+|\det K_1|\big)^2.
\end{equation}

Since $C^2_{ES}\geq 0$, it follows from Eqs. (\ref{SS'Ebis}) that $d_S>0$ and $d_E>0$ are sufficient conditions for having, respectively, $C^2_{S|S'E}>0$ and $C^2_{E|SS'}>0$ (of course, the condition $C^2_{ES}> 0$ also suffices for guaranteeing entanglement in both bipartitions, yet it involves the specific form of the initial state $\rho_0$ ---via the dependence of $C^2_{ES}$ on $G$--- and would not lead to a condition involving the Kraus operators only). This leads to
\begin{subequations}\label{SyE}
\begin{eqnarray}
0<|\det K_0|+|\det K_1|&\Rightarrow& C^2_{S|S'E}>0,\label{Sa}\\
|\det K_0|+|\det K_1|<1&\Rightarrow& C^2_{E|SS'}>0,\label{Ea}
\end{eqnarray}
\end{subequations}
and consequently to the final condition (of course, provided $\tau=0$):
\begin{eqnarray}\label{SufW}
0<|\det K_0|+|\det K_1|<1 
\Rightarrow \ket{\phi(p)} \textrm{has W-type}\\ \textrm{tripartite entanglement}\nonumber.
\end{eqnarray}
It is worth noting that since (\ref{SufW}) ensued from imposing that the terms proportional to $\mathcal{E}^2_0$ in Eqs. (\ref{SS'Ebis}) were strictly positive, it constitutes a condition for having \emph{distributed} W-type genuine entanglement, i.e., for guaranteeing the emergence of tripartite entanglement given that an initial amount of initial (bipartite) entanglement was present. By imposing instead the less restrictive conditions $C^2_{S|S'E}>0$ and $C^2_{E|SS'}>0$ in Eqs. (\ref{SS'Ebis}) we would obtained more general conditions for $\ket{\phi(p)}$ to exhibit W-type entanglement; however, as mentioned before and by virtue of the term $C^2_{ES}$, those conditions will in general depend on the initial state $\rho_0$. Thus, in general, the presence of W-type entanglement depends on the initial state $\rho_0$, contrary to the 3-tangle (\ref{FenDet}), which is independent of $\rho_0$.  

The above results serve also to establish the conditions under which $K_0$ and $K_1$ drive the initial state to a biseparable one, i.e., separable in either the bipartition $S|S'E$ or $E|SS'$. Specifically, from Eqs. (\ref{SEtau0}), (\ref{SS'Ebis}) and (\ref{Db}) we arrive at
\begin{subequations}\label{bisep}
\begin{align}
C^2_{S|S'E}&=0\iff\det K_0=\det K_1=G=0\\
& \quad\quad\quad\Rightarrow C^2_{E|SS'}=\mathcal{E}^2_0,\nonumber\\
C^2_{E|SS'}&=0\iff G=\mathcal{E}^2_0\big(|\det K_0|^2+|\det K_1|^2-1\big)\nonumber\\
& \quad\quad\quad\Rightarrow C^2_{S|S'E}=\mathcal{E}^2_0.
\end{align}
\end{subequations}
The first of these equations thus determines the conditions for having a state that is $S$-separable (yet entangled in the $E+S'$ subsystem), whereas the second one corresponds to an $E$-separable state (in which all the entanglement is between $S$ and $S'$). 

A summary of the results obtained in this section is shown in Table \ref{T}, where the different inequivalent families and each type of entanglement are identified according to the conditions on the Kraus operators. Notice that such conditions are invariant under (local) unitary transformations $U_S$, hence do not depend on the basis in which the matrices are expressed. 
\begin{center}
\begin{table}
\begin{tabular}{ |c|c|c|}
\hline
\emph{Family} & \emph{Condition} & \emph{Entanglement}\\
\hline \hline

GHZ & $4\det (K_{0}K_{1})\neq g^{2}(K_{0},K_{1})$   & 3-tangle\\
\hline\hline
W &  $4\det (K_{0}K_{1})= g^{2}(K_{0},K_{1})$ & no 3-tangle\\ 
\hline\hline
& $0<|\det K_0|+|\det K_1|<1$ & 3-partite \\ \cline{2-3}
W-&$\det K_0=\det K_1=$& biseparable\\
subfamilies&$=G(K,\rho_0)=0$& ($S$-separable)\\\cline{2-3}
&  $G(K,\rho_0)=$& biseparable\\
&  $=\mathcal{E}^2_0\big(|\det K_0|^2+|\det K_1|^2-1\big)$& ($E$-separable)\\
\hline 
\end{tabular}
\caption{Families to which the evolved state $\ket{\phi(p)}$ pertains according to basic properties of the Kraus operators. All conditions are sufficient and necessary, except for the condition involving the W-subfamily of 3-partite entangled states, which is only sufficient. Note that $S'$-separable and fully separable states do not occur since we are considering $C^2_{S'|SE}=\mathcal{E}^2_0>0$.}\label{T}
\end{table}
\end{center}
\subsection{Tripartite entanglement as a function of the initial entanglement}\label{ad1}
The results shown in Table \ref{T} were obtained assuming a nonzero initial (bipartite) entanglement $\mathcal E^2_0$. As follows from the previous analysis, it is clear that the condition $\mathcal E^2_0>0$ is necessary in order to create tripartite entanglement (otherwise the only entanglement present would be, at most, $C^2_{SE}$), yet the dependence of the generation of tripartite entanglement on the amount of initial entanglement has not been discussed. We now briefly comment on this.

Equation (\ref{FenDet}) shows that the 3-tangle depends linearly on $\mathcal E^2_0$, so for an appropriate given evolution (fixed Kraus operators that comply with the conditions (\ref{iff})), an increase/decrease of the initial entanglement changes the GHZ-type tripartite entanglement accordingly, and 3-tangle exists for all $\mathcal E^2_0\in(0,1]$. 

The analysis for the W-type tripartite entanglement is more elaborated, since the expressions for $C^2_{i|jk}$ in Eqs. (\ref{SS'E}) are not explicit functions of $\mathcal E^2_0$, due to the term $G(K,\rho_0)$ and its dependence on the initial state $\rho_0$, whose determinant gives precisely $\mathcal E^2_0$ (see Eq. (\ref{E0detrhoB})). In order to express $G$ as an explicit function of the initial entanglement, we first write $\rho_0$ in the form 
\beq \label{rhoM}
\rho_0=\rho_0(\rho_{ee}, \varphi, |\rho_{ge}|)=\left(\begin{array}{cc}
1-\rho_{ee} & |\rho_{ge}|e^{i\varphi}  \\
|\rho_{ge}|e^{-i\varphi} & \rho_{ee}  \\
 \end{array} \right),
\eeq  
and decompose the matrix $K_{0}\rho_0K^{\dagger}_{1}$ as
\begin{equation}\label{MN}
K_{0}\rho_0K^{\dagger}_{1}=M(K,\rho_{ee})+|\rho_{ge}|N(K,\varphi),
\end{equation}
with $M$ and $N$ the matrices 
\begin{equation}\label{MNel}
M=K_{0}\sigma_{ee}K^{\dagger}_{1},\quad N=K_{0}\sigma_{\varphi}K^{\dagger}_{1},
\end{equation}
and
\beq \label{sigmas}
\sigma_{ee}=\left(\begin{array}{cc}
1-\rho_{ee} & 0  \\
0 & \rho_{ee}  \\
 \end{array} \right),\quad \sigma_{\varphi}=\left(\begin{array}{cc}
0 & e^{i\varphi}  \\
e^{-i\varphi} & 0  \\
 \end{array} \right).
\eeq  
Substituting Eq. (\ref{MN}) into (\ref{G}) we obtain, with the aid of the properties (\ref{propsg}),
\begin{gather}\label{GMN1}
G(K,\rho_0)=-4\,g(M,M^{\dagger})-\nonumber\\
-8|\rho_{ge}|\,\textrm{Re}\,[g(M,N^{\dagger})]-4|\rho_{ge}|^2\,g(N,N^{\dagger}).
\end{gather}
Now, resorting to Eq. (\ref{E0detrhoB}) written explicitly as  
\begin{equation} \label{cond}
 \mathcal{E}^2_0=4\rho_{ee}(1-\rho_{ee})-4|\rho_{eg}|^2,
\end{equation}
we can write $|\rho_{ge}|$ as a function of the initial excited population $\rho_{ee}$ and entanglement $\mathcal{E}^2_0$; the set of the independent variables $(\rho_{ee}, \varphi, |\rho_{ge}|)$ becomes substituted by $(\rho_{ee}, \varphi, \mathcal{E}^2_0)$, and we are led to  
\begin{gather}
G(K_0,\rho_0)=G(K, \rho_{ee}, \varphi,\mathcal{E}^2_0)=\nonumber\\
=\mathcal{G}_1(K,\rho_{ee},\varphi)+\mathcal{G}_2(K,\rho_{ee},\varphi)\sqrt{4\rho_{ee}(1-\rho_{ee})-\mathcal{E}^2_0}+\nonumber\\
+\,\mathcal{G}_3(K,\varphi)\,\mathcal{E}^2_0,\label{GMNbis}
\end{gather}
with
\begin{subequations}
\begin{eqnarray}
\mathcal{G}_1&=&\!-4\,[g(M,M^{\dagger})+\rho_{ee}(1-\rho_{ee})g(N,N^{\dagger})],\\
\mathcal{G}_2&=&\!-4\,\textrm{Re}\,[g(M,N^{\dagger})],\\
\mathcal{G}_3&=&g(N,N^{\dagger}).
\end{eqnarray}
\end{subequations}

When we substitute Eq. (\ref{GMNbis}) into (\ref{SEtau0}) we get, from Eqs. (\ref{SS'Ebis}), 
\begin{gather}
\frac{\partial C^2_{i|jS'}}{\partial \mathcal{E}^2_0}\!=(d_i+2|\det(K_0K_1)|)+\nonumber\\
+\,g(N,N^{\dagger})+\frac{2\,\textrm{Re}\,[g(M,N^{\dagger})]}{\sqrt{4\rho_{ee}(1-\rho_{ee})-\mathcal{E}^2_0}},\label{partial}
\end{gather}
where $i,j=S,E$ (and of course $i\neq j$). Equation (\ref{partial}) gives the general expression for the change of the bipartite entanglements $C^2_{S|S'E}$ and $C^2_{E|SS'}$ with respect to $\mathcal{E}^2_0$ in absence of 3-tangle, thus allows to analyze the behaviour of bipartite and W-type tripartite entanglement as $\mathcal{E}^2_0$ varies. The specific rate of change will clearly depend on the set of Kraus operators, and on the value of $\mathcal{E}^2_0$ provided $\textrm{Re}\,[g(M,N^{\dagger})]\neq 0$. However, it should be stressed that the conditions for the emergence of bipartite or W-type genuine entanglement summarized in Table \ref{T} do not depend on the amount of initial entanglement, so whenever those conditions are satisfied, the corresponding type of entanglement exists for all $\mathcal E^2_0\in(0,1]$.   
\\
\subsection{Emergence of multipartite entanglement in 4-qubit systems}
The approach developed here throws some light into the emergence of multipartite entanglement in 4-qubit systems in which two of them ($S$ and $S'$), initially entangled, interact locally with the remaining two ($E$ and $E'$). The initial state is thus of the form
\begin{equation}\label{4OriginalState}
\ket{\phi(0)}_{S'SEE'} = \ket{\psi_0}_{S'S}\ket{0}_{E}\ket{0}_{E'},
\end{equation}
where $\ket{\psi_0}_{S'S}$ is an entangled state, and the evolution operator factorizes as $U=U'_{S'E'}\otimes U_{SE}$, with $U'_{S'E'}$ and $U_{SE}$ arbitrary unitary operators associated, respectively, to the set of Kraus operators $\{K'_0,K'_1\}$ and $\{K_0,K_1\}$. The dynamics of entanglement in this type of systems has been studied in \cite{17}, where it is shown that the residual entanglement in the bipartition $i|jkl$, namely
\begin{equation}\label{Ridef}
R_i\equiv C^2_{i|jkl}-C^2_{ij}-C^2_{ik}-C^2_{il},
\end{equation}
can be expressed as
\begin{equation}\label{Ri}
R_i= \tau_{\underline{i}kl}+\tau_{ij(kl)}  
\end{equation}
for $i,j\in\{S,E\}$ and $k,l\in\{S',E'\}$, or $i,j\in\{S',E'\}$ and $k,l\in\{S,E\}$. Here $\tau_{\underline{i}kl}$ stands for the residual entanglement ---corresponding to the bipartition $i|kl$--- of the (in general mixed) reduced 3-qubit state $\rho_{ikl}=\textrm{Tr}_{j}\rho$, i.e.,
\begin{equation}\label{taui}
\tau_{\underline{i}kl}\equiv C^2_{i|kl}-C^2_{ik}-C^2_{il},  
\end{equation}
and $\tau_{ij(kl)}$ represents the 3-tangle of the pure evolved state $U\ket{\phi(0)}$,  generated among the subsystems $i$, $j$ and $k+l$ (as discussed in \cite{17}, the subsystem $k+l$ behaves as a two-level system, hence can be considered as an effective qubit, here denoted as $(kl)$). Equation (\ref{Ri}) thus shows that the residual entanglement can be decomposed in terms of well-identified (physically and operationally) multipartite entanglement contributions involving 3 and 4 qubits (see \cite{17} for details). 

Now, due to the invariance of entanglement under local operations, the quantities $C^2_{S|ES'E'}$, $C^2_{E|SS'E'}$, $C^2_{SE}$, $C^2_{S|S'E'}$, $C^2_{E|S'E'}$, and $\tau_{SE(S'E')}$ do not depend on the particular transformation $U'_{S'E'}$, and consequently can be computed setting $U'_{S'E'}=\mathbb I_{S'E'}$. In this case the qubit $E'$ becomes ineffective (remains disentangled from the rest), and formally the problem amounts to solve the 3-qubit system studied here. This observation leads thus to
\begin{subequations}\label{4y3}
\begin{eqnarray}
C^2_{S|ES'E'}&=&C^2_{S|S'E}(K_0,K_1,\rho_0),\label{64a}\\
C^2_{E|SS'E'}&=&C^2_{E|SS'}(K_0,K_1,\rho_0),\label{64b}\\
C^2_{SE}&=&C^2_{SE}(K_0,K_1,\rho_0),\\
C^2_{S|S'E'}&=&C^2_{SS'}(K_0,K_1,\rho_0),\\
C^2_{E|S'E'}&=&C^2_{ES'}(K_0,K_1,\rho_0),\\
\tau_{SE(S'E')}&=&\tau_{SS'E}(K_0,K_1),\label{tau64}
\end{eqnarray}
\end{subequations}
where the left-hand side expressions refer to the 4-qubit problem, and the quantities in the right-hand side correspond to those derived in the previous sections, for the 3-qubit case. It is important to stress that the equalities in Eqs. (\ref{4y3}) represent only a numerical (not physical) correspondence, and that they hold for \emph{all} $U=U'_{S'E'}\otimes U_{SE}$ (considering $U'_{S'E'}=\mathbb I_{S'E'}$ was only a useful assumption that served to establish the numerical equivalence). Analogously, we can make the substitution $i\leftrightarrow i'$ ($i=S,S',E,E'$) throughout the previous analysis (including Eqs. (\ref{SS'E}) and (\ref{bip})), and proceed mutatis mutandis to obtain: 
\begin{subequations}\label{4y3b}
\begin{eqnarray}
C^2_{S'|E'SE}&=&C^2_{S'|E'S}(K'_0,K'_1,\rho'_0),\label{65a}\\
C^2_{E'|S'SE}&=&C^2_{E'|S'S}(K'_0,K'_1,\rho'_0),\label{65b}\\
C^2_{S'E'}&=&C^2_{S'E'}(K'_0,K'_1,\rho'_0),\\
C^2_{S'|SE}&=&C^2_{SS'}(K'_0,K'_1,\rho'_0),\\
C^2_{E'|SE}&=&C^2_{E'S}(K'_0,K'_1,\rho'_0),\\
\tau_{S'E'(SE)}&=&\tau_{SS'E'}(K'_0,K'_1).\label{tau65}
\end{eqnarray}
\end{subequations}

From Table \ref{T} and Eqs. (\ref{tau64}), (\ref{tau65}), and (\ref{Ri}), it follows that whenever the Kraus operators corresponding to the transformation $U_{SE}$, and those corresponding to the transformation $U'_{S'E'}$ comply with the conditions
\begin{subequations}\label{condKK'}
\begin{eqnarray}
4\det(K_0K_1)&\neq& g^2(K_0,K_1),\\
4\det(K'_0K'_1)&\neq& g^2(K'_0,K'_1),
\end{eqnarray}
\end{subequations}
all $R_i$ ($i=S, E, S',E'$) will be nonzero, and therefore multipartite entanglement (embodied in the residual entanglement) will be present in the 4-qubit system. Notice that conditions (\ref{condKK'}) guarantee only multipartite entanglement in the form of $\tau_{ij(kl)}$. In order to guarantee tripartite entanglement in the form $\tau_{\underline{i}kl}$, the explicit expressions for the qubit-qubit entanglements $C^2_{ik}$, $C^2_{il}$ (with one primed index and one unprimed index) need to be calculated directly, as functions of the Kraus operators. 

 Also, it follows from Table \ref{T} and Eqs. (\ref{64a}), (\ref{64b}), (\ref{65a}), and (\ref{65b}), that whenever $U_{SE}$ and $U'_{S'E'}$ generate W-type tripartite entanglement, that is, whenever
\begin{subequations}
\begin{gather}\label{condKK'W}
4\det(K_0K_1)=g^2(K_0,K_1),\;0\!<\!|\!\det K_0|\!+\! |\!\det K_1|\!<\!1,\\
4\det(K'_0K'_1)=g^2(K'_0,K'_1),\;0\!<\!|\!\det K'_0|\!+\! |\!\det K'_1|\!<\!1,
\end{gather}
\end{subequations}
 the 4-qubit system will be entangled in all bipartitions, thus guaranteeing the emergence of genuine entanglement. 
\\
\section{Examples}\label{ex}
We will now apply the results obtained for the 3-qubit system to analyze the distribution and emergence of genuine entanglement, considering different quantum channels of interest and an initial state $\rho_{0}$ expressed as (\ref{rhoM}).

\subsection{Amplitude Damping Channel}

\noindent The Amplitude Damping (AD) Channel represents a dissipative interaction between $S$ and $E$. A paradigmatic example is that of an initially excited two-level atom ($S$) that decays spontaneously in an initially empty cavity ($E$), which absorbs the emitted photon with some probability $p$ \cite{Aud}. The corresponding Kraus operators are:

 \begin{equation}
  K_0 = \left( \begin{array}{cc}
1 & 0  \\
0 & \sqrt{1-p}  \\
 \end{array} \right), \quad K_1 = \left( \begin{array}{cc}
0 & \sqrt{p}  \\
0 & 0 \\
 \end{array} \right),\label{ADC}
 \end{equation}
with $p\in[0,1]$. 

Direct calculation gives $u=4\det(K_0K_1)=0$, $v=g^2(K_0,K_1)=0$ for all $p$, whence $\tau=0$ and any initial state (\ref{OriginalState}) that evolves under the AD map pertains to the W-family. Moreover, since $\det K_1=0$ and $\det K_0=\sqrt{1-p}>0$ for $p<1$, condition (\ref{SufW}) becomes
\begin{eqnarray}\label{SufWAD}
\!0<p<1 
\Rightarrow \ket{\phi(p)} \textrm{is (W-type) genuine entangled}.
\end{eqnarray}
We also get $G=4\rho_{ee}^2p(1-p)$, so the condition for having a $S$-separable state (see Table \ref{T}) is satisfied only at $p=1$. Analogously, the condition for having an $E$-separable state holds only at $p=0$. This simple analysis on the Kraus operators thus leads us to conclude that under the AD map, W-type genuine entanglement is created during all the evolution, except at the extreme points $p=0,1$, in which the entanglement exists only in bipartite form. At $p=0$ there is only entanglement between $S$ and $S'$, and at $p=1$ the same amount of entanglement is completely transferred between $E$ and $S'$. 

\begin{figure}
\begin{center}
\includegraphics[scale=0.4,angle=0]{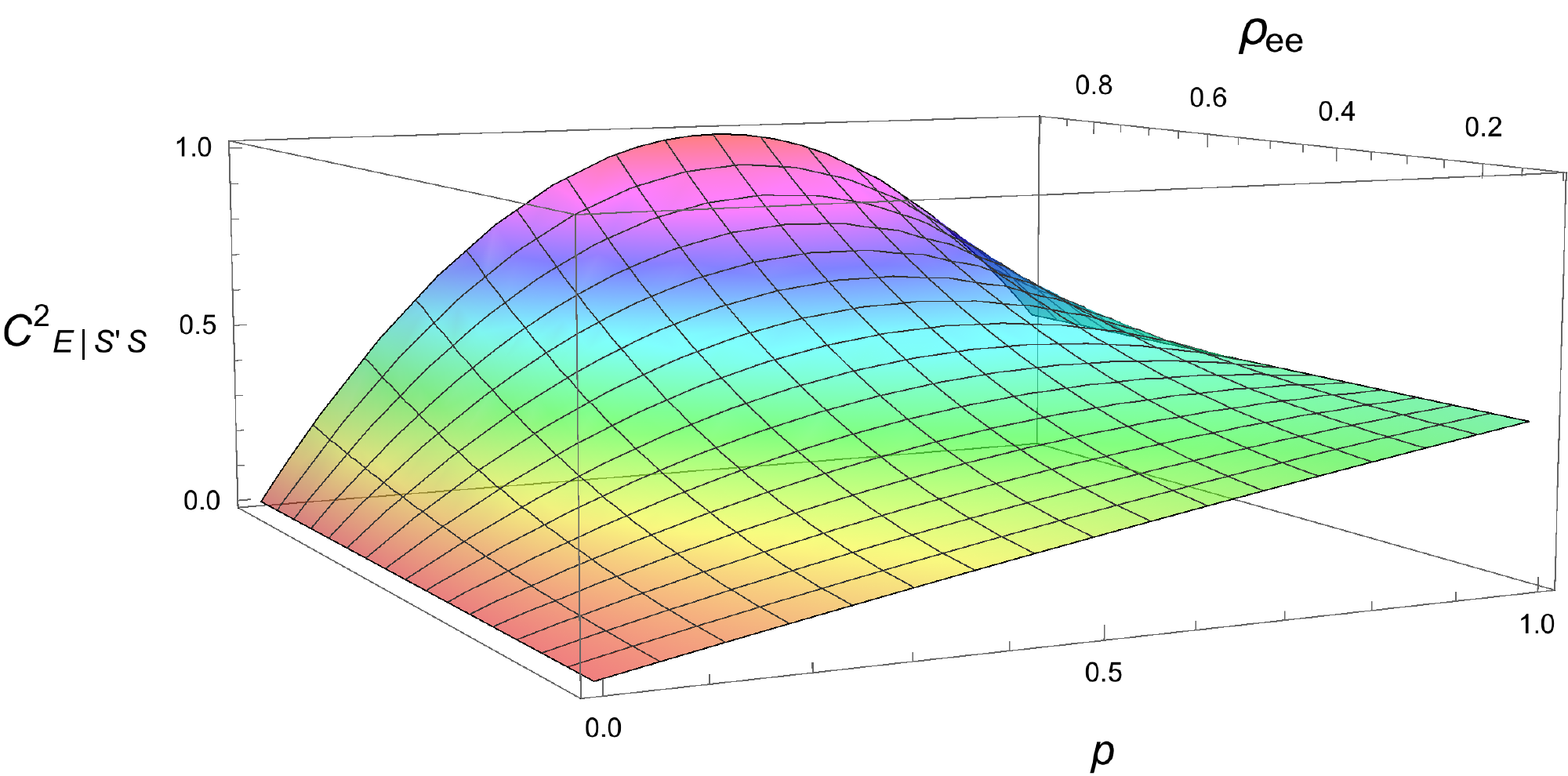}
\caption{$C^{2}_{E|SS'}$ as a function of $p$ and $\rho_{ee}$, for $\mathcal{E}^2_0=0.4$ under the AD channel. Subsystem $E$ is initially ($p=0$) disentangled from the rest, it gets entangled with the $S+S'$ system during the evolution, and at $p=1$ becomes entangled with $S'$ only. W-type genuine entanglement exists for all $0<p<1$.    
 \label{F1}}
 \includegraphics[scale=0.4,angle=0]{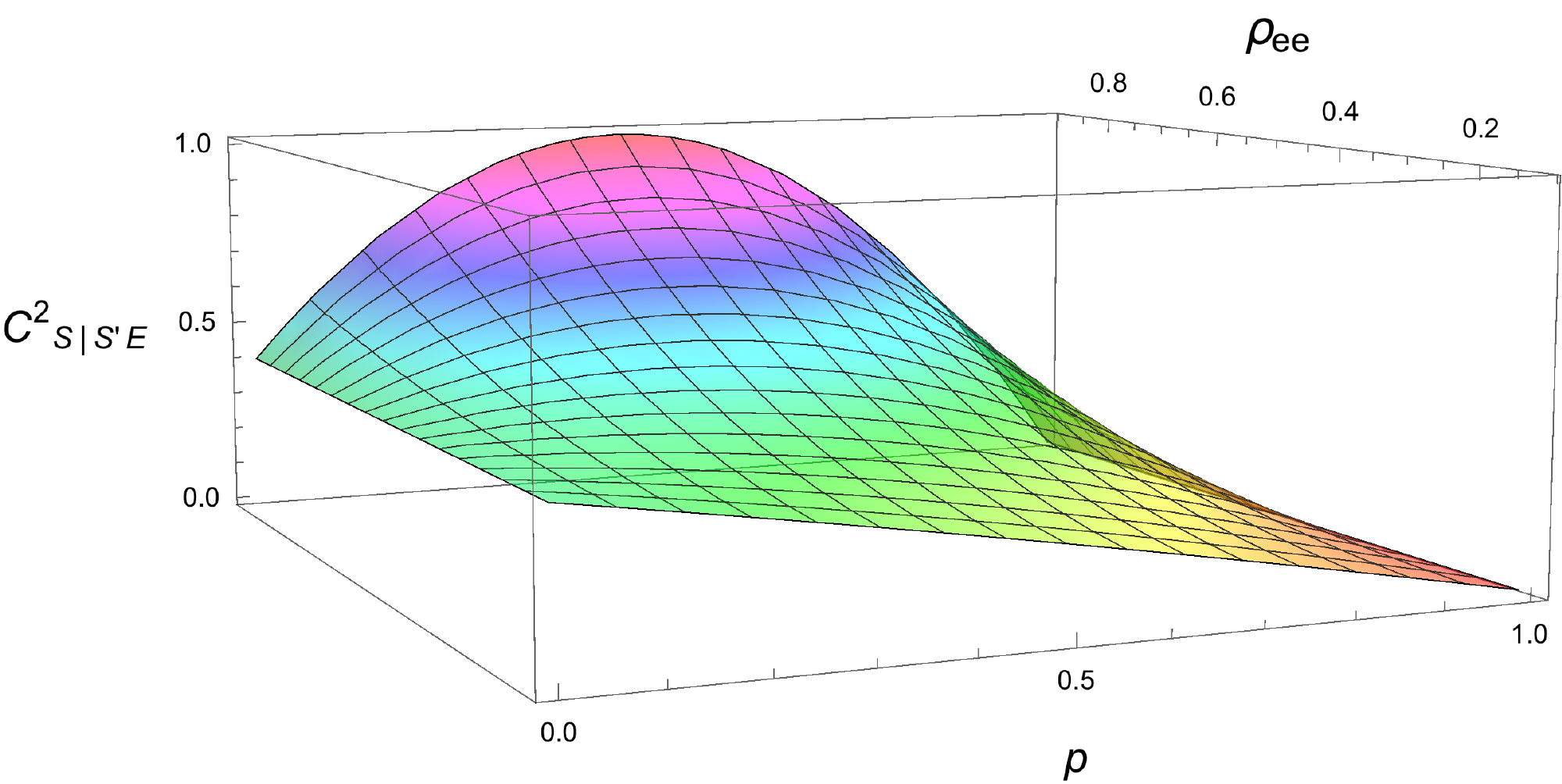}
\caption{$C^{2}_{S|S'E}$ as a function of $p$ and $\rho_{ee}$, for $\mathcal{E}^2_0=0.4$ under the AD channel. Subsystem $S$ is initially ($p=0$) entangled with $S'$, its entanglement with the $E+S'$ system starts to decrease or increase depending on the value of $\rho_{ee}$ (see text), and finally $S$ disentangles from the rest, at $p=1$. W-type genuine entanglement exists for all $0<p<1$.
 \label{F2}}
 \includegraphics[scale=0.4,angle=0]{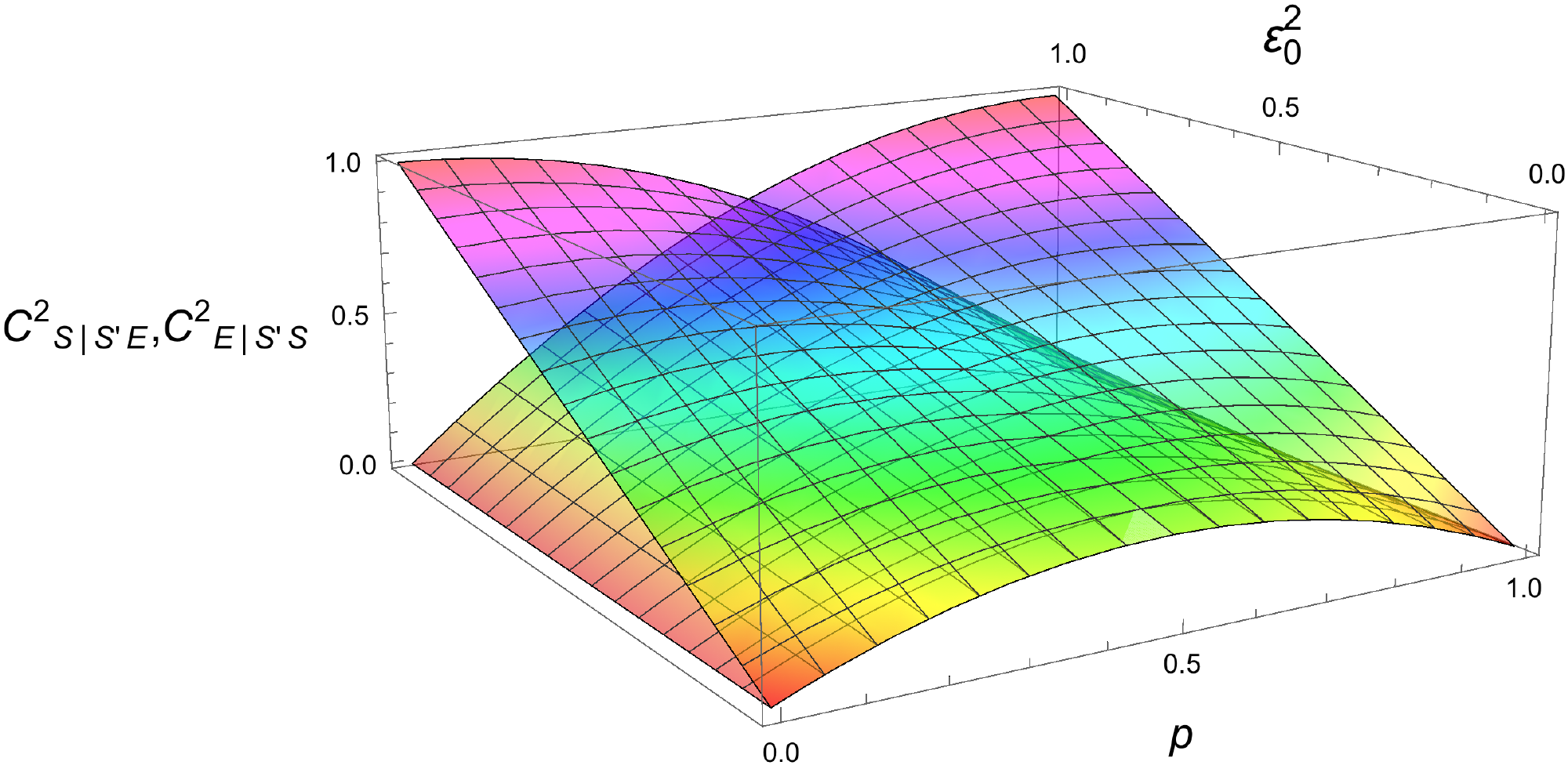}
\caption{$C^{2}_{S|S'E}$ and $C^{2}_{E|SS'}$ as a function of $p$ and $\mathcal{E}^2_0$, for $\rho_{ee}=0.5$ under the AD channel. $C^{2}_{S|S'E}$  ($C^{2}_{E|SS'}$) decreases (increases) as $p\rightarrow 1$, and entanglement in all bipartitions --hence W-type tripartite entanglement-- exists for all $p\in(0,1)$ and $\mathcal{E}^2_0\in(0,1]$.     
 \label{F1N}}
\end{center}
\end{figure}

To analyze the distribution of entanglement we resort to the following expressions, obtained by direct application of Eqs. (\ref{bip2}): 
\begin{subequations}\label{bipAD}
\begin{align}
C^{2}_{S'S}&=\mathcal{E}^2_0(1-p), \label{SS'AD}
\\
C^{2}_{S'E}&=\mathcal{E}^2_0p,\label{S'EAD}
\\
C^{2}_{SE}&=G=4\rho_{ee}^2p(1-p).\label{S'EAD}
\end{align}    
\end{subequations}
Further, Eqs. (\ref{SS'Ebis}) lead to
\begin{subequations}\label{todosAD}
\begin{align}
C^{2}_{S|S'E}&=\mathcal{E}^2_0(1-p)+4\rho_{ee}^2p(1-p), \label{SAD}
\\
C^{2}_{E|SS'}&=\mathcal{E}^2_0p+4\rho_{ee}^2p(1-p).\label{EAD}
\end{align}
\end{subequations}
Figures (\ref{F1}) and (\ref{F2}) show the evolution of these entanglements as a function of $p$ and $\rho_{ee}$, for $\mathcal{E}^2_0=0.4$. Notice that though $\rho_{ee}$ can acquire values in the interval $[0,1]$, Eq. (\ref{cond}) imposes $\mathcal{E}^2_0-4\rho_{ee}(1-\rho_{ee})\leq 0$, so for $\mathcal{E}^2_0$ fixed, the range of $\rho_{ee}$ is restricted to $\rho_{ee}\in [\rho^{-}_{ee},\rho^{+}_{ee}]$, where $\rho^{\pm}_{ee}$ are the roots of the equation $\mathcal{E}^2_0-4\rho_{ee}(1-\rho_{ee})=0$, namely $\rho^{\pm}_{ee}=(1/2)(1\pm \sqrt{1-\mathcal{E}^2_0})$. 

As seen in Figs. (\ref{F1}) and (\ref{F2}), for fixed $p\in(0,1)$, the entanglement in both bipartitions $S|S'E$ and $E|SS'$ increases monotonically with the initial population of excited states in $S$. Also, as $\rho_{ee}\rightarrow 0$, $C^{2}_{E|SS'}$ and $C^{2}_{S|ES'}$ increases and decreases, respectively, linearly with $p$ and at the same rate. Notice that once the evolution starts ($p=0$) the entanglement $C^{2}_{E|SS'}$ always increases (irrespective of the value of $\rho_{ee}$), yet the sign of $(\partial C^{2}_{S|ES'}/\partial p)|_{p=0}$ depends on 
the value of $\rho_{ee}$: for $\rho_{ee}<\mathcal{E}_0/2$ the initial entanglement in the bipartition $S|ES'$ starts to decrease, but increases for $\rho_{ee}>\mathcal{E}_0/2$. 

Figure (\ref{F1N}) shows the entanglement in the bipartitions $E|SS'$ and $S|S'E$ as a function of $p$ and $\mathcal{E}^2_0$, for $\rho_{ee}=0.5$. It serves to illustrate the presence of W-type entanglement as the amount of initial entanglement varies. In line with (\ref{SufWAD}), we verify that entanglement exists in all bipartitions for $0<p<1$, irrespective of the value of $\mathcal{E}^2_0$.
\subsection{Dephasing Channel}
We now analyze the evolution under the Dephasing (D) Channel, which has the following Kraus operators:
 \begin{equation}
  K_0 = \left( \begin{array}{cc}
1 & 0  \\
0 & \sqrt{1-p}  \\
 \end{array} \right), \quad K_1 = \left( \begin{array}{cc}
0 & 0  \\
0 & \sqrt{p} \\
 \end{array} \right),\label{DC}
 \end{equation}
again with $p\in[0,1]$. This channel represents a non-dissipative interaction between $S$ and $E$, describing, e.g., an elastic scattering with probability $p$, where the state of $S$ does not change, but $E$ is allowed to perform a transition without exchanging energy with $S$ \cite{Aud}. 

Direct calculation gives $u=4\det(K_0K_1)=0$, so that $\tau =\mathcal{E}_0^2|g^2(K_0,K_1)|=\mathcal{E}_0^2p$, and thus
\begin{eqnarray}\label{tauDD}
\!\!0\!<p\leq1 
\Rightarrow \! \ket{\phi(p)} \textrm{is\! (GHZ-type)\! genuine entangled}.
\end{eqnarray}

We now resort to Eqs. (\ref{bip}) to calculate all qubit-qubit entanglements, obtaining 
\begin{subequations}\label{bipDC}
\begin{align}
C^2_{S'S} &= \mathcal{E}^2_0(1-p),\\
C^2_{S'E} &= 0,    \\
C^2_{SE} &= G=4p|\rho_{eg}|^2.
\end{align}
\end{subequations}
The above results lead to
\begin{subequations}\label{todosDC}
\begin{eqnarray}
C^2_{S|S'E}&=&\mathcal{E}^2_0+4p|\rho_{eg}|^2\nonumber\\
&=&\mathcal{E}^2_0(1-p)+4\rho_{ee}(1-\rho_{ee})p,\\
C^2_{E|SS'}&=&\mathcal{E}^2_0p+4p|\rho_{eg}|^2\nonumber\\
&=&	4\rho_{ee}(1-\rho_{ee})p,                   
\end{eqnarray}
\end{subequations}
where in the last line Eq. (\ref{cond}) has been used. 
\begin{figure}
\begin{center}
\includegraphics[scale=0.4,angle=0]{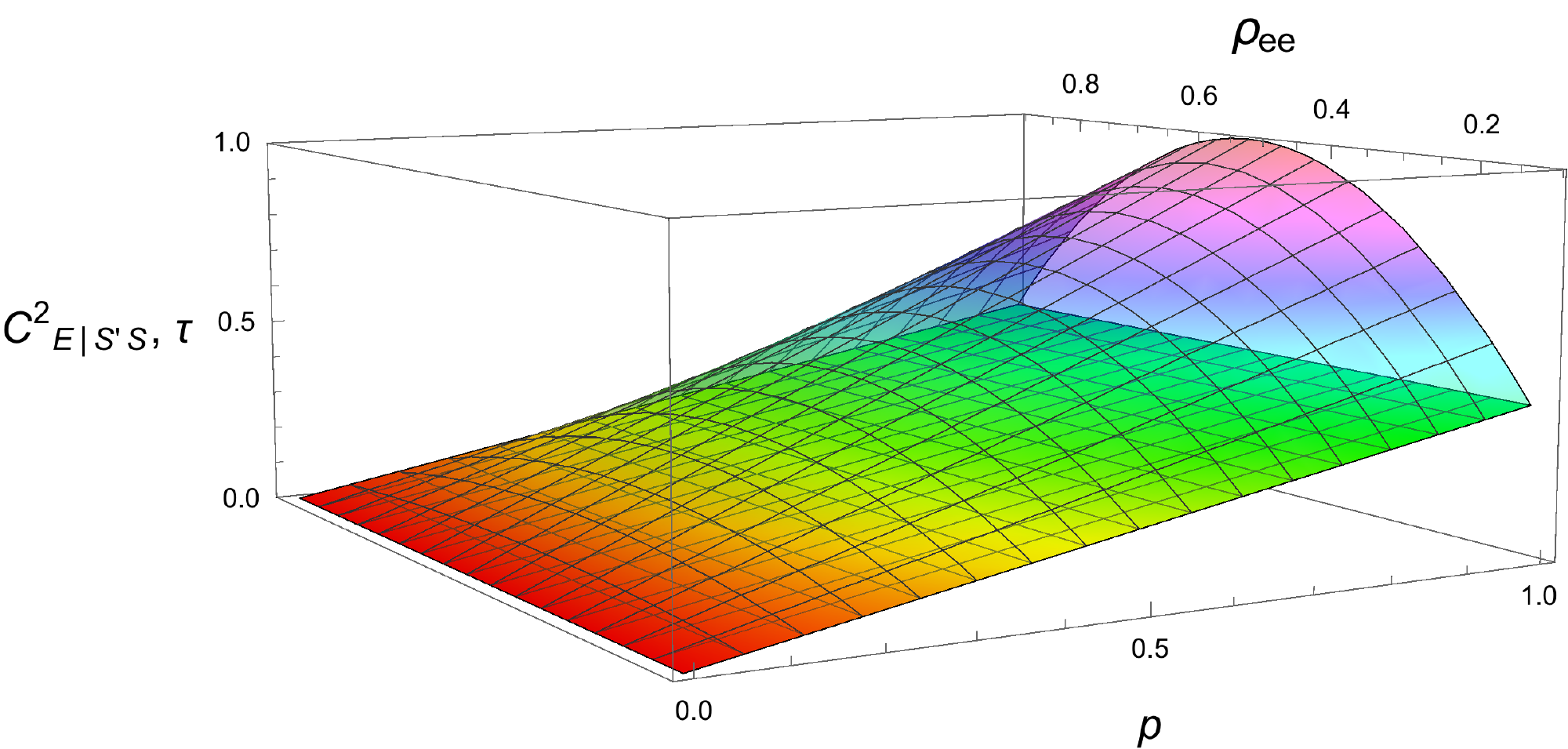}
\caption{$C^{2}_{E|SS'}$ and $\tau$ as a function of $p$ and $\rho_{ee}$, for $\mathcal{E}^2_0=0.4$ under the D channel. Subsystem $E$ is initially disentangled from the rest. As the system evolves $C^{2}_{E|SS'}$ (curved surface) increases linearly with $p$ with a slope that depends on $\rho_{ee}$, whereas $\tau$ (flat surface) increases linearly with $p$, at a rate independent of $\rho_{ee}$. At $p=1$ and $\rho_{ee}=1/2$, $C^{2}_{E|SS'}$ acquires its maximum value.     
 \label{F3}}
 \includegraphics[scale=0.4,angle=0]{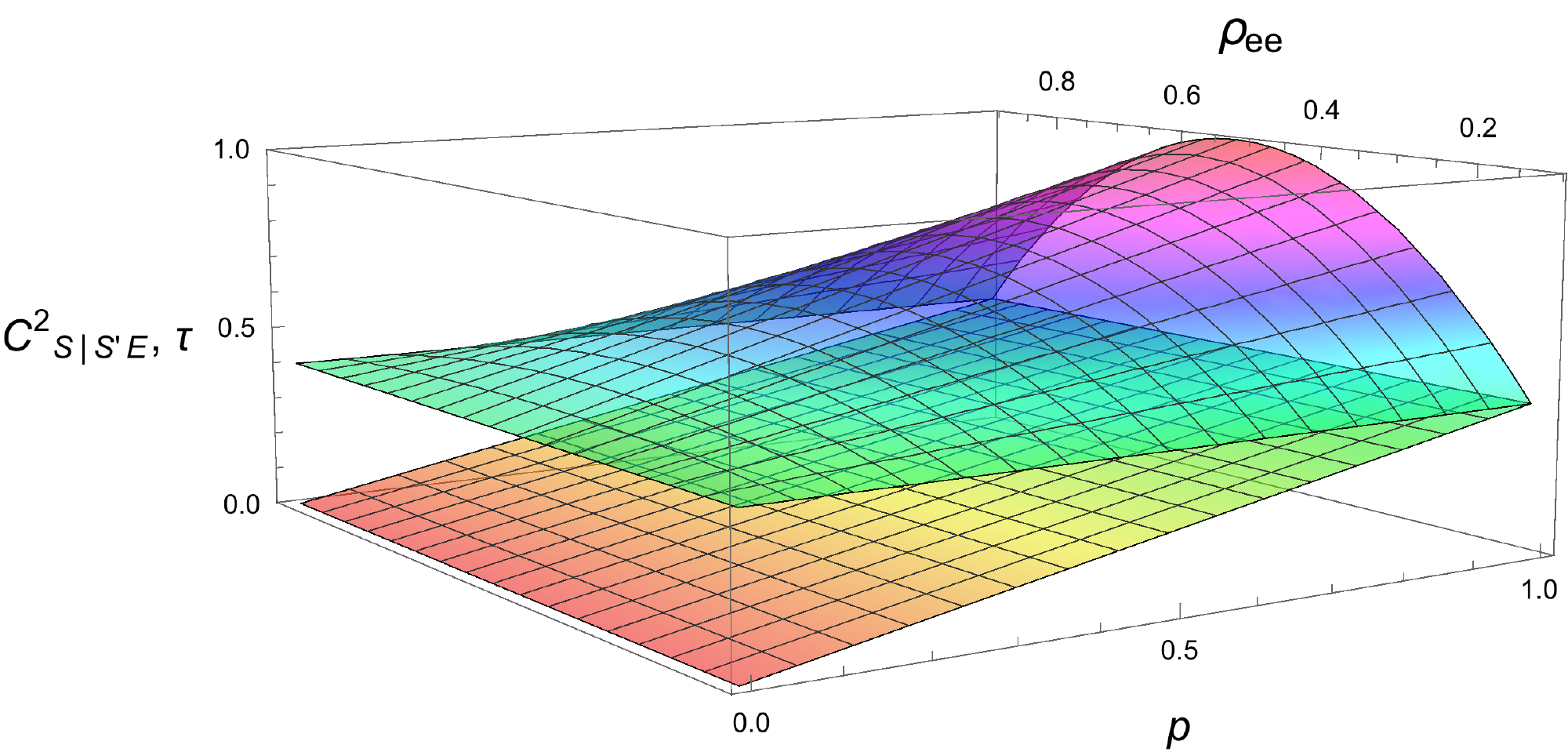}
\caption{$C^{2}_{S|S'E}$ and $\tau$ as a function of $p$ and $\rho_{ee}$, for $\mathcal{E}^2_0=0.4$ under the D channel. At $p=0$ all the entanglement exists in bipartite form (between $S$ and $S'$). During the evolution, $C^{2}_{S|S'E}$ (curved surface) increases at a constant rate that depends on $\rho_{ee}$, and $\tau$ (flat surface) increases linearly with $p$. At $p=1$ and $\rho_{ee}=1/2$, $C^{2}_{S|S'E}$ acquires its maximum value.    
 \label{F4}}
 \includegraphics[scale=0.4,angle=0]{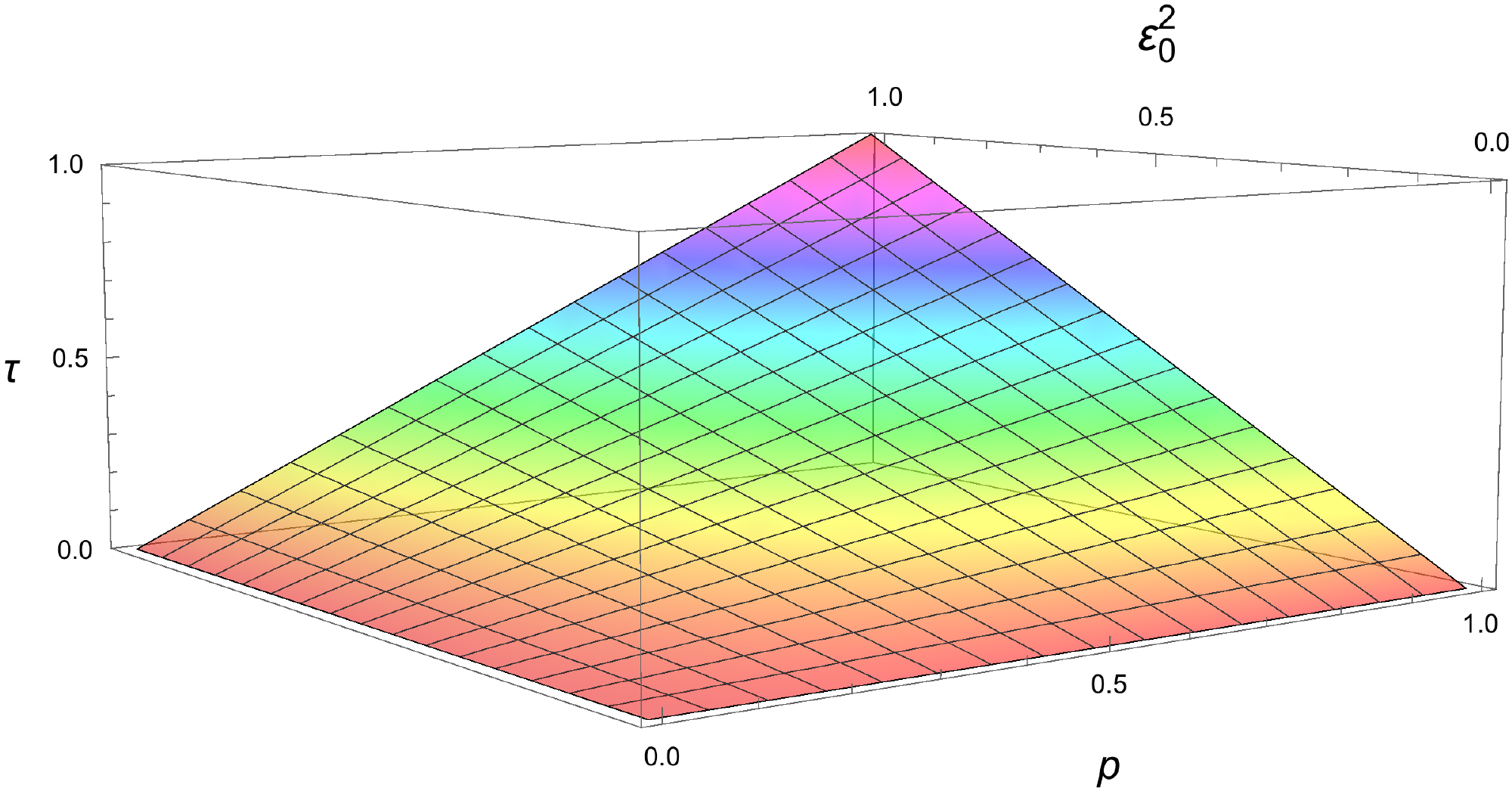}
\caption{$\tau$ as a function of $p$ and $\mathcal{E}^2_0$ for $\rho_{ee}=0.5$ under the D channel. The GHZ-type genuine entanglement increases with $\mathcal{E}^2_0$.    
 \label{F3N}}
\end{center}
\end{figure}

Figures (\ref{F3}) and (\ref{F4}) show the evolution of $C^2_{E|SS'}$, $C^2_{S|S'E}$, and $\tau$ as functions of $p\in[0,1]$ and $\rho_{ee}\in [\rho^{-}_{ee},\rho^{+}_{ee}]$, for $\mathcal{E}^2_0=0.4$. We observe that
under the dephasing evolution the initial entanglement distributes in such a way that the 3-tangle increases linearly and attains its maximum value at $p=1$, where the only nonzero qubit-qubit entanglement is $C^2_{SE}(p=1)=4|\rho_{eg}|^2$. This means that for $\rho_{eg}=0$ (i.e., $\rho_{ee}=\rho^{\pm}_{ee}$), \emph{all} the amount of initial bipartite entanglement $\mathcal{E}^2_0$ is completely transformed (at $p=1$) into genuine (GHZ-type) entanglement.

Figure (\ref{F3N}) shows the variation of the 3-tangle as a function of $p$ and $\mathcal{E}^2_0$ for $\rho_{ee}=0.5$. As expected, the GHZ-genuine entanglement increases with the amount of initial entanglement.

\subsection{Phase Flip Channel}

The Phase Flip (PF) Channel corresponds to one of the possible errors that can occur, with probability  $p/2$, in quantum computation \cite{Aud}. It is characterized by the following Kraus operators:

\begin{equation}
  K_0 =\sqrt{1-p/2} \left( \begin{array}{cc}
1 & 0  \\
0 & 1  \\
 \end{array} \right), K_1 = \sqrt{p/2} \left( \begin{array}{cc}
1 & 0  \\
0 & -1 \\
 \end{array}\right),
 \end{equation}
with $p \in [0,1]$. In this case we get $u=4\det(K_0K_1)=p(p-2)<0$ for all positive $p$, which is a sufficient condition for having nonzero 3-tangle (see below Eq. (\ref{iff})). Hence 
\begin{eqnarray}\label{tauPF}
\!\!0\!<p\leq1 
\Rightarrow \! \ket{\phi(p)} \textrm{is\! (GHZ-type)\! genuine entangled},
\end{eqnarray}
as in the dephasing channel case. However, here we have $v=g^2(K_0,K_1)$$=0$, and consequently from Eqs. (\ref{FenDet}) and (\ref{bip}) we find 
\begin{subequations}
\begin{align}
\tau&=\mathcal{E}^2_0 p(2-p),\\
C^2_{SS'} &= \mathcal{E}^2_0(1-p)^2,\\
C^2_{S'E} &= 0,    \\
C^2_{SE} &= \big[4\rho_{ee}(1-\rho_{ee})-\mathcal{E}^2_0\big]p(2-p).
\end{align}
\end{subequations}
Further, we get
\begin{subequations}\label{PF}
\begin{align}
C^2_{S|S'E}&=\mathcal{E}^2_0(1-p)^2 + 4\rho_{ee}(1-\rho_{ee})p(2-p),\\
C^2_{E|SS'}&=4\rho_{ee}(1-\rho_{ee})p(2-p).	                   
\end{align}
\end{subequations}
\begin{figure}
\begin{center}
\includegraphics[scale=0.4,angle=0]{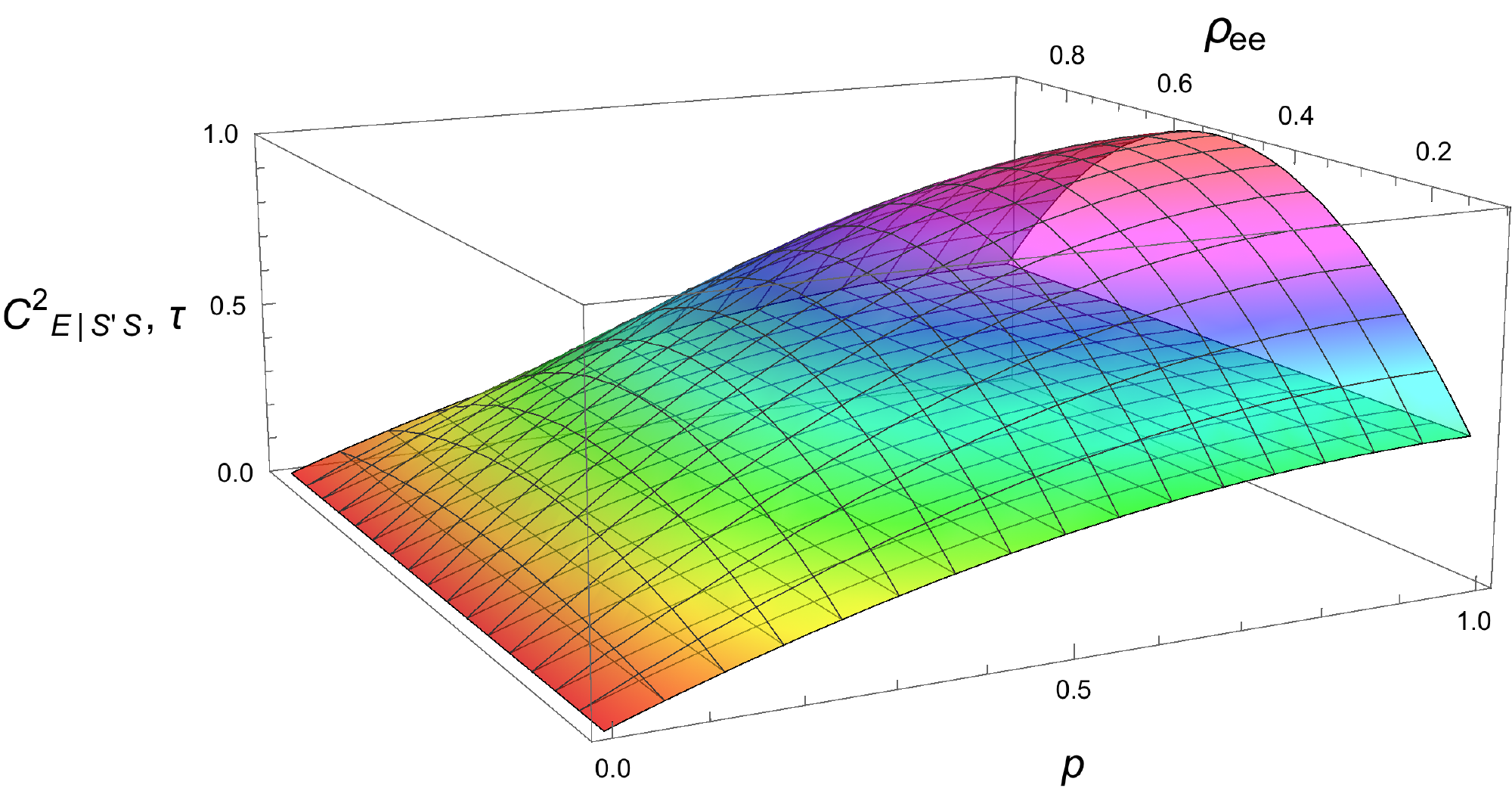}
\caption{$C^{2}_{E|SS'}$ and $\tau$ as a function of $p$ and $\rho_{ee}$, for $\mathcal{E}^2_0=0.4$ under the PF map.  Along the evolution, $C^{2}_{E|SS'}$ (outer surface) and $\tau$ (inner surface) increase monotonically with $p$, the latter in a way that is independent of the initial excited population $\rho_{ee}$. At $p=1$ the curves coincide with those of the D map. \label{F5}}
\includegraphics[scale=0.4,angle=0]{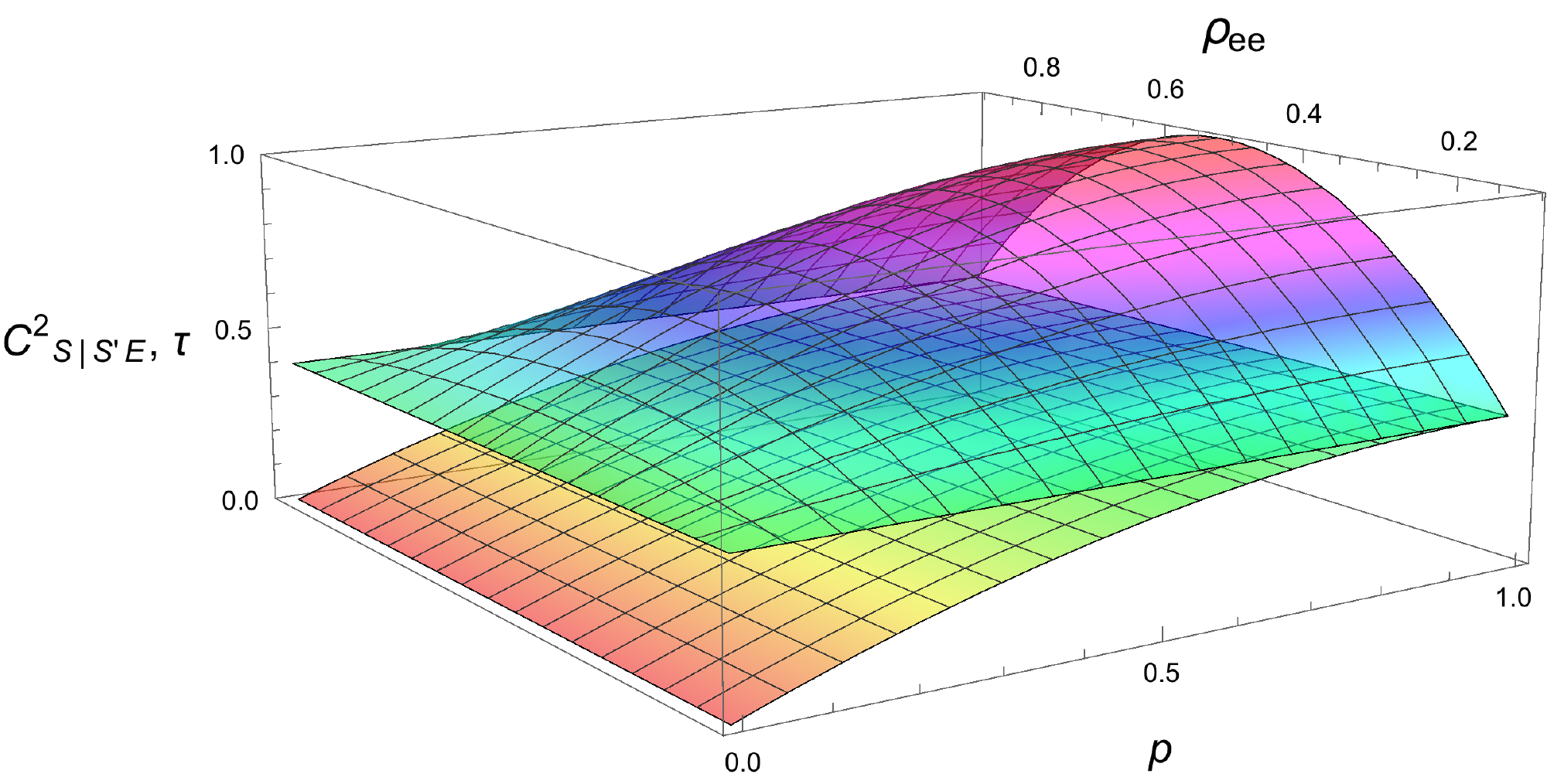}
\caption{$C^{2}_{S|S'E}$ and $\tau$ as a function of $p$ and $\rho_{ee}$, for $\mathcal{E}^2_0=0.4$ under the PF map. During the evolution, $C^{2}_{S|S'E}$ (outer surface) and $\tau$ (inner surface) increase monotonically with $p$, yet the behaviour of the 3-tangle does not depend on $\rho_{ee}$. At the final point ($p=1$) the curves coincide with those of the D map. \label{F6}}
\includegraphics[scale=0.4,angle=0]{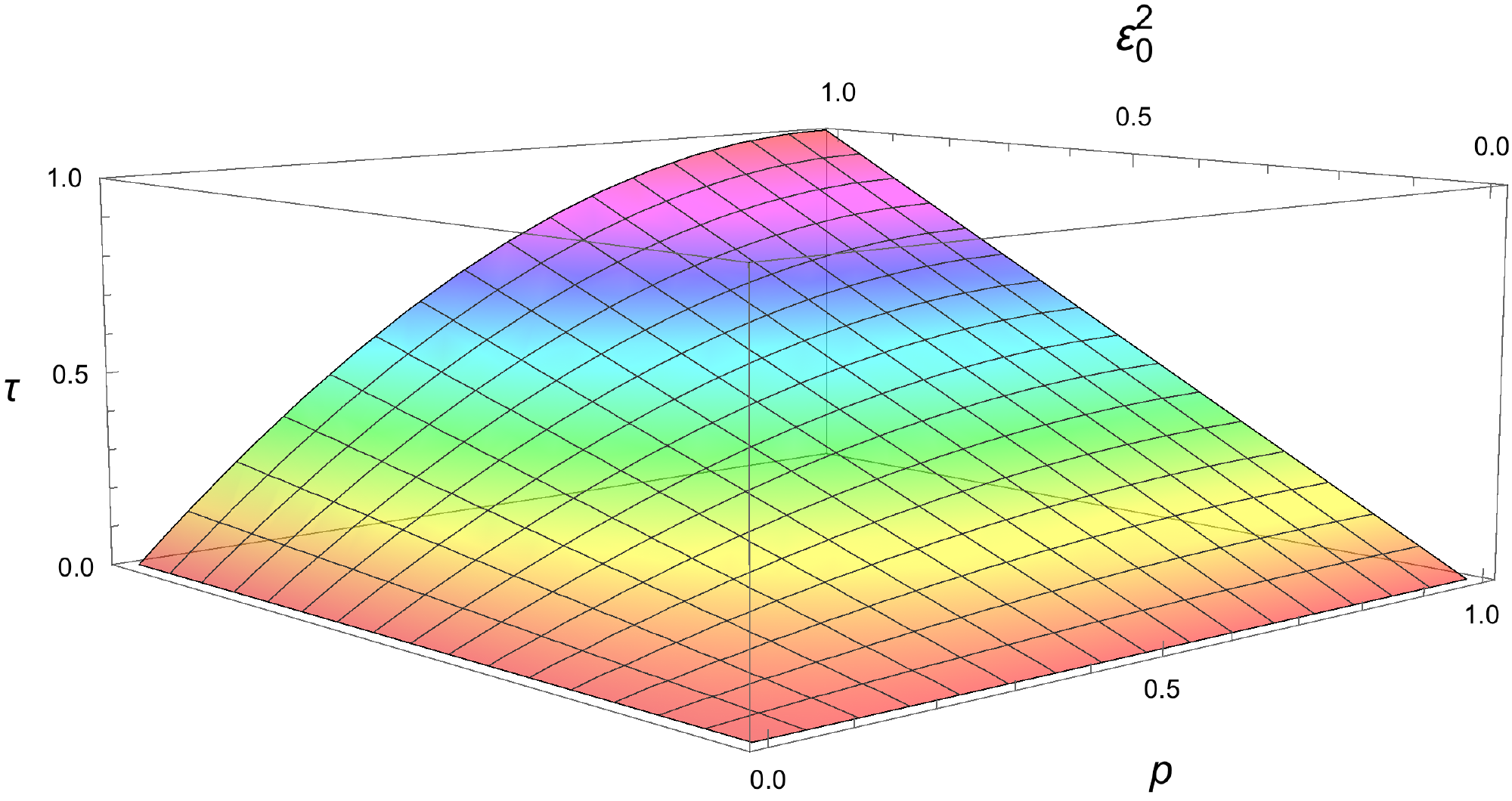}
\caption{$\tau$ as a function of $p$ and $\mathcal{E}^2_0$ for $\rho_{ee}=0.5$ under the PF map. The GHZ-type genuine entanglement increases with $\mathcal{E}^2_0$. \label{F5N}}
\end{center}
\end{figure}
Figures (\ref{F5}) and (\ref{F6}) show, respectively, the evolution of $C^2_{E|SS'}$ and $C^2_{S|S'E}$ given by Eqs. (\ref{PF}), together with $\tau$ as a function of $p\in[0,1]$ and $\rho_{ee}\in [\rho^{-}_{ee},\rho^{+}_{ee}]$, for $\mathcal{E}^2_0=0.4$. The dynamics of these quantities is similar to that exhibited in the D channel case, but now all entanglements evolve quadratically (instead of linearly) in $p$. 

Finally, Figure (\ref{F5N}) shows the emergence of 3-tangle as a result of the evolution and the nonzero initial entanglement. The GHZ-type genuine entanglement is alway larger that in the Dephasing channel case.
\section{Concluding remarks}\label{conclu}

We disclosed basic and unitarily-invariant properties that the Kraus operators must satisfy in order to guarantee the emergence of each type of entanglement allowed in a 3-qubit pure state, assuming an evolution dictated by Eq. (\ref{rho}). We found that the condition distinguishing states in the GHZ-family from those in the W-family depends \emph{only} on functions involving determinants of the Kraus operators, which means that the emergence of 3-tangle is determined by the structure of the evolution operator only, and is independent of the particular initial (entangled) state $\rho_0$. Sufficient conditions for having W-type genuine entanglement are also found that are independent of the initial state. Conditions involving $\rho_0$ appear instead in the criteria for having biseparability.  

Our analysis provides the expression for \emph{all} entanglement measures, namely $C^2_{ij}, C^2_{i|jk}$, and $\tau$, in terms of the Kraus operators and a single function ($G$) that isolates all the information regarding the initial state $\rho_0$, and that determines the entanglement between $S$ and $E$. Further, a general expression for $G$ as a function of the initial entanglement was obtained, thus allowing for a more complete analysis of the different types of entanglement as $\mathcal{E}^2_0$ varies. In addition, lower bounds are found for the qubit-qubit entanglements. 

Since we have considered evolutions in which $S$ is initially entangled with $S'$, and the only interaction is between $S$ and $E$, our analysis refers to 3-partite entanglement \emph{induced by local channels, and assisted by the initial entanglement}. If we allow both $S$ \emph{and} $S'$ to interact with $E$, the dynamics would be described by a \emph{global} channel, communication between $S$ and $S'$ through $E$ will arise in general, and the emergence of all kinds of entanglement is possible ---even if $S$ and $S'$ are initially uncorrelated--- since the evolution corresponds to nonlocal dynamics.   

Our results allow for a detailed analysis of the dynamics, emergence and distribution of entanglement in a wide range of scenarios, particularly in those involving decoherence processes, as those depicted in the examples. In all these, it became clear that the loss of entanglement between the central systems (in this case $S$ and $S'$) is accompanied by the creation of genuine entanglement among $S$, $S'$, and $E$ (which in this case plays the role of environment). If access to the degrees of freedom of $E$ were possible (as in, e.g., \cite{15}) decoherence could therefore be used to create or enhance useful genuine entanglement, by mere application of the appropriate Kraus operators. 

The present classification of the Kraus operators according to their capacity of producing specific types of entanglement, opens the path to investigate whether similar analysis can be performed on larger systems (such as the 4-qubit case briefly studied here), and also to determine the basic structure of the evolution operators ---or rather of the specific Hamiltonians--- that must be implemented in the system in order to create and distribute quantum correlations as required for specific tasks. This would favor the development and experimental implementation of strategies aimed at making the most of initially entangled, evolving systems.   
\acknowledgments{The authors gratefully acknowledge financial support from DGAPA-UNAM through Project PAPIIT
IA101918. They also thank A. Leonides for helping in the elaboration of the graphical abstract.}
{}
\end{document}